\documentclass[a4paper,10pt]{article}
\usepackage[utf8]{inputenc}
\usepackage[left=2.5cm,right=2.5cm,top=1.5cm,bottom=2.5cm,footnotesep=0.5cm]{geometry}
\PassOptionsToPackage{hyphens}{url}
\usepackage[hidelinks]{hyperref}

\usepackage[square,numbers]{natbib}
\usepackage{amsmath}
\usepackage{amsfonts}
\usepackage{amssymb}
\usepackage{graphicx}
\usepackage{subcaption}
\usepackage{comment}
\usepackage[skip=1pt]{caption}
\usepackage{fancyhdr}
\usepackage[normalem]{ulem}
\usepackage{longtable}
\usepackage{amsthm}
\usepackage{amsmath}
\usepackage{pifont}
\usepackage{lscape}
\usepackage{color}
\usepackage{tabularray}
\usepackage{array}
\usepackage{tikz}
\usepackage{colortbl}
\usepackage{authblk}
\usepackage{caption}
\usepackage[hyphens]{url}
\usepackage[hyphenbreaks]{breakurl}

\usepackage[format=hang,labelfont=bf]{caption}

\theoremstyle{definition}
\title{A Review of Cybersecurity Incidents in the \\Food and Agriculture Sector}

\author[1]{Ajay Kulkarni}
\author[2]{Yingjie Wang}
\author[3]{Munisamy Gopinath}
\author[4]{Dan Sobien}
\author[5]{Abdul Rahman}
\author[1,2,6]{Feras A. Batarseh}

\date{}

\affil[1]{Commonwealth Cyber Initiative, Virginia Tech, Arlington, VA 22203, USA}
\affil[2]{Bradley Department of Electrical and Computer Engineering, Virginia Tech, Arlington, VA 22203, USA}
\affil[3]{Department of
Agricultural and Applied Economics, University of Georgia, Athens, GA 30602, USA}
\affil[4]{National Security Institute, Virginia Tech, Arlington, VA 22203, USA}
\affil[5]{Deloitte \& Touche LLP, Baltimore, MD 21202, USA}
\affil[6]{Department of Biological Systems Engineering, Virginia Tech, Blacksburg, VA 24060, USA}

\begin{document}
\maketitle
\fancyhf{}
\renewcommand{\headrulewidth}{0pt}
\renewcommand{\footrulewidth}{0.4pt}
\pagenumbering{gobble}

\normalsize
\pagenumbering{arabic}
\subsection*{Abstract\footnotetext[1]{ajaysk@vt.edu}
\footnotetext[2]{chelseawang@vt.edu}
\footnotetext[3]{m.gopinath@uga.edu}
\footnotetext[4]{sdan8@vt.edu}
\footnotetext[5]{abdulrahman@deloitte.com}
\footnotetext[6]{batarseh@vt.edu}}

\noindent The increasing utilization of emerging technologies in the Food \& Agriculture (FA) sector has heightened the need for security to minimize cyber risks. Considering this aspect, this manuscript reviews disclosed and documented cybersecurity incidents in the FA sector. For this purpose, thirty cybersecurity incidents were identified, which took place between July 2011 and April 2023. The details of these incidents are reported from multiple sources such as: the private industry and flash notifications generated by the Federal Bureau of Investigation (FBI), internal reports from the affected organizations, and available media sources. Considering the available information, a brief description of the security threat, ransom amount, and impact on the organization are discussed for each incident. This review reports an increased frequency of cybersecurity threats to the FA sector. To minimize these cyber risks, popular cybersecurity frameworks and recent agriculture-specific cybersecurity solutions are also discussed. Further, the need for AI assurance in the FA sector is explained, and the Farmer-Centered AI (FCAI) framework is proposed. The main aim of the FCAI framework is to support farmers in decision-making for agricultural production, by incorporating AI assurance. Lastly, the effects of the reported cyber incidents on other critical infrastructures, food security, and the economy are noted, along with specifying the open issues for future development.

\section{Introduction}

Critical infrastructures are ``the assets, systems, networks, facilities, and other elements that society relies upon to maintain national security, economic vitality, and public health, and safety"
~\cite{cisa2019guide}. The U.S. Department of Homeland Security (DHS) categorized the 16 critical infrastructure sectors as follows: 1) Chemical, 2) Commercial Facilities, 3) Communications, 4) Critical Manufacturing, 5) Dams, 6) Defense Industrial Base, 7) Emergency Services, 8) Energy, 9) Financial Services, 10) Food and Agriculture, 11) Government Facilities, 12) Healthcare and Public Health, 13) Information Technology, 14) Nuclear Reactors, Materials, and Waste, 15) Transportation Systems, and 16) Water and Wastewater Systems~\cite{cisa2019guide}. The elements of these 16 infrastructure sectors have connections and interdependencies~\cite{petit2015analysis}, and losing one or more lifeline functions immediately impacts the others. For this survey study, the focus is on the Food and Agriculture (FA) sector. In 2021, the U.S. Department of Agriculture (USDA)\cite{usdaUSDAFood} reported that agriculture and related industries have a 5.4\% share (\$1.264 trillion) of the U.S. Gross Domestic Product (GDP), and farms contribute about 0.7\% (\$164.7) of the U.S. GDP. Additionally, the FA sector provided 10.5\% (21.1 million) of the U.S. employment in 2021, and an average American household spends 12\% of their household's budget on food~\cite{usdaUSDAFood}. Based on the Food and Agriculture Organization (FAO) of the United Nations (UN), the global demand for food will grow to 70\% to meet the population by 2050~\cite{hunter2017agriculture}. This underscores the importance of the FA sector, making it one of the nation's critical infrastructures.

The current advances, especially considering the fourth industrial revolution (Industry 4.0), are responsible for the utilization of emerging technologies such as sensors, the Internet of Things (IoT), Artificial Intelligence (AI), big data, embedded systems, communication networks, and robotics in the FA domain~\cite{liu2020industry}. It is deemed possible to increase the quantity and quality of agriculture and food products using these emerging technologies in the farm management cycle ~\cite{wolfert2017big}. This new age of agriculture is refered to as Agriculture 4.0~\cite{zhang2002precision}. A smart farm is a cyber-physical system whose operations can be monitored and controlled by a computer and communication system using a set of networked agents~\cite{barreto2017industry,navarro2020systematic}. These network agents can be sensors, linear actuators, control processing units, and communication devices~\cite{sontowski2020cyber}. As noted by Zanella et al.~\cite{de2020security}, smart farming utilizes emerging technologies, and due to their complexity, it incorporates all the security problems present in these technologies into smart farms, making them vulnerable to security threats. Additionally, the U.S. Food \& Drug Administration (FDA)~\cite{fdaHACCPPrinciples} endorses a Hazard Analysis Critical Control Point (HACCP) management system for ensuring food safety from harvest to consumption. Nowadays, many food processors utilize automated HACCP systems, which, if compromised, could cause food-borne illness outbreaks affecting food security and the economy. Chalk~\cite{chalk2001terrorism} reported that a significant attack on the FA sector could result in mass economic destabilization, including financial losses, lack of food security, loss of political support \& confidence in the government, and social instability. Additionally, the FA sector has received less attention than other sectors, especially regarding accurate threat assessments, response structures, and preparedness initiatives~\cite{chalk2001terrorism,barreto2018smart}. These reasons make it crucial to protect the FA sector from attacks, which can be physical and/or cyber~\cite{archivesReportCost}.

\subsection{Motivation} 

Leveraging sensor technologies can make network layers (Perception, Network, Edge, and Application) more vulnerable to cyberattacks, and different types of cyberattacks occur on them are presented in Table~\ref{tblr:attacktab}~\cite{de2020security}. 
Snotowski et al.~\cite{sontowski2020cyber} also reported various kinds of network attacks, such as password cracking, evil twin access points, key reinstallation, and spoofing that can happen on Agriculture 4.0 infrastructures. Yazdinejad et al.~\cite{yazdinejad2021review} noted that privacy, integrity, confidentiality, availability of the services, non-repudiation, and trust are affected by cyberattacks at the farm as well. The FBI also has provided warnings~\cite{FBIa,FBIb,FBIc,FBId,FBIe} on ransomware attacks on the FA sector and noted the growing impact of these attacks. Based on FBI notifications, the attackers are gradually broadening their cyber activity by attacking information technology, business processes, and operational technologies assets. The DHS~\cite{boghossian2018threats} also reported multiple cybersecurity scenarios threatening Agriculture 4.0, including:

\begin{itemize}
\itemsep0em 
\item A malicious actor alters data or algorithms in livestock or thoroughbred management software in a competitor’s breeding stock, causing them to miss breeding gestation windows. 

\item An attacker ingests rough data into smart farm equipment sensor networks during the planting and harvesting season, affecting automated agricultural decision support systems and food security.

\item Water supply facility uses sensors to control the level of chemicals in water; in cases where the data used for decision-making is compromised, the water gets poisoned, affecting irrigation systems.

\item A foreign company dominates the commercial Unmanned Aerial Vehicles (UAV) market configured to store sensor data in a foreign country or installs built-in firmware for remote access tools.

\item A farming co-op faces data compromise, including sensitive economic and personal data, or compromised USDA policy and economic datasets.
\end{itemize}

\begin{figure}[h]
\centering
\includegraphics[width=\textwidth]{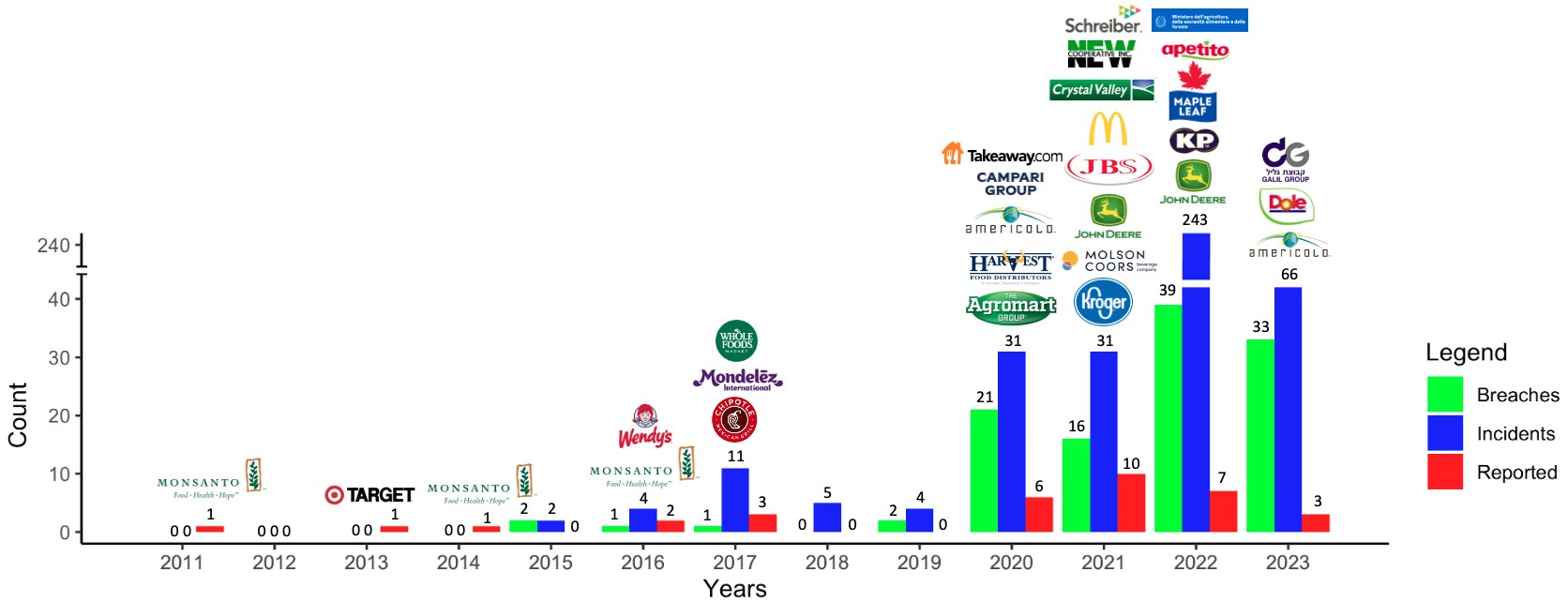}
\captionsetup{format=hang}
\caption{The timeline of 397 security incidents and 115 security breaches from 2015 to 2023 in the FA domain. The security incidents reported in this survey are shown using company logos from 2011 to 2023.}
\label{fig:figure1}
\end{figure}

\noindent In addition to the listed scenarios, Yazdinejad et al.~\cite{yazdinejad2021review} noted that attackers can use different attacks on FA infrastructure, such as ransomware, data leakage, and false data injection on data that is being stored or processed in the database. Further, end-user applications like decision-support tools can be attacked to harm or destroy normal operations of Agriculture 4.0~\cite{yazdinejad2021review}. The annual data breach investigation reports published by Verizon~\cite{Verizon} indicated 397 security incidents and 115 security breaches between 2015 and 2023, only in the FA domain. The report differentiates between the two kinds of incidents, according to Verizon~\cite{Verizon}, a security incident is “a security event that compromises the integrity, confidentiality or availability of an information asset”, and a security breach is “an incident that results in the confirmed disclosure — not just potential exposure — of data to an unauthorized party”. Based on these reports, the highest number of security threats are noted for 2022, and yearly details are visualized in Figure~\ref{fig:figure1}. 

The increasing number of cybersecurity threats makes it essential to detect and mitigate them along with improving situational awareness to enhance resiliency at the farm~\cite{fda2022Food}. Considering these aspects, compiling a comprehensive list of incidents that occurred in the domain would help in evaluating and improving the readiness and resilience of the FA sector against such attacks. 
Further, their effects on critical infrastructures, food security, and the economy are essential to consider for strategizing preparedness, response, and recovery in the FA domain. Ordinary citizens can be impacted due to these heterogeneous effects, also causing an impact on the FA domain globally. Accordingly, this paper's main contributions include:

\begin{itemize}
\itemsep0em 
    \item A compilation and review of the 30 cybersecurity incidents reported in the FA sector between 2011 and 2023.
    
    \item Reporting of popular cybersecurity frameworks and recent agriculture-specific cybersecurity solutions.

    \item Proposing and describing a novel Farmer-Centered AI (FCAI) framework that incorporates AI assurance in agriculture applications, which is a need of the hour as highlighted by DHS~\cite{cisaSoftwareMust}.
    
    \item Discussing the effects of cybersecurity incidents on other critical infrastructures, food security, and the economy.

\end{itemize}

\NewTblrTheme{fancy}{  
\SetTblrStyle{caption-tag}{font=\bfseries, black} 
\SetTblrStyle{caption-sep}{font=\bfseries, black} 
}

\begin{longtblr}[
    theme = fancy,
  caption = {Network layers, resources, and different attacks in Agriculture 4.0. The details on resources and attacks are based on Zanella et al.~\cite{de2020security}.},
  label = {tblr:attacktab}
]{
  width = \linewidth,
  colspec = {Q[112]Q[385]Q[442]},
  row{1} = {c},
  row{2} = {c},
  row{3} = {c},
  row{4} = {c},
  row{5} = {c},
  vlines,
  hline{1-2} = {1-3}{},
  hline{3-6} = {-}{},
}
\textbf{Network Layer} & \textbf{Resources} & \textbf{Attacks}\\
Perception & Senors, Camera, Actuator, GPS, Tag RFID & {Random sensor incidents, Autonomous system hijacking and disruption, Optical deformation, Irregular measurements, Sensor weakening, Node capture, Fake node, Sleep deprivation}\\
Network & Router, Access points, Switches, Protocols, other resources & {Denial of Service (DoS)/Distributed DoS (DDoS), Data transit attacks, Routing attacks, Signal disruptions}\\
Edge & Security features, Gateway, In-out interface, Diverse resources & {Forged control for actuators, Gateway-cloud request forgery, Forged measure injection, Booting, Unauthorised access, Man-in-the-middle, Signature wrapping, Flooding}\\
Application & Database, Web tool, Decision-making, End-user application & Phishing, Malicious scripts, DoS/DDoS
\end{longtblr}

The next section presents the need for this survey and the inclusion-exclusion criteria for the survey's collection methodology.
    
\subsection{The Need for this Survey}

A rising interest in the research community is witnessed 
to provide novel solutions for securing the FA sector~\cite{sharma2020machine,huerta2021wireless,unal2020smart}. Additionally, recent surveys~\cite{liu2020industry, de2020security,kayan2022cybersecurity,rudrakar2023iot} on smart agriculture discuss details of security challenges and provide a general review of available cyber methods in literature.  However, the available literature needs to provide a comprehensive compilation and review of the actual cybersecurity incidents in the FA sector. This is essential because it can assist in learning the sheer diversity of cyberattacks and their consequences while designing mitigation strategies. For instance, DoS/DDoS attacks can halt the data transfer between a farm and a food producer, affecting the farmer's decision-making. Similarly, ransomware attacks can interrupt daily agricultural operations such as mixing fertilizers and logistical operations, or halting operations in food processing plants.
Thus, a review of the actual cybersecurity incidents can assist in making better cyber policies, strategies, and defense solutions across the board. Recently, Hassanzadeh et al.~\cite{hassanzadeh2020review} have documented fifteen actual cybersecurity incidents in the Water and Wastewater sector. These have been used as a reference by the Department of Economic and Social Affairs (DESA) at the UN to provide recommendations on the digitalization of water systems~\cite{barrie2022recommendations}. Also, reporting on these incidents has been used in cybersecurity policy and the legislative contexts~\cite{malatji2020cybersecurity}. Considering these details and to bridge a gap between actual cybersecurity incidents and cyber defense solutions for agriculture, we have compiled them in this survey, which is an essential distinction between this survey paper and other existing surveys. In this survey, we use the term \textit{``agriculture"} encompassingly by including the food and retail sectors. 

The following are the inclusion-exclusion criteria for our survey's collection methodology. The selection of cybersecurity incidents in this survey is based on a definition provided by Hassanzadeh et al.~\cite{hassanzadeh2020review} and refers to ``an incident that has been maliciously launched from cyberspace to cause adverse consequences to a target entity". Considering this definition, this survey includes only those cybersecurity incidents that are identified, disclosed, and documented in the food, agriculture, and retail sectors between July 2011 and April 2023. These inclusion-exclusion criteria resulted in 34 cybersecurity incidents, and then the categorization of these incidents is per organization, which provided us with 30 cybersecurity incidents. For example, three Data Breach incidents occurred at Monsanto (July 2011, March 2014 \& January 2016), but they are grouped under ``Monsanto". Similarly, Americold Logistics faced two cyberattacks, one in November 2020 and one in April 2023. These cyberattacks are also grouped under ``Americold Logistics". 
The sources for the reported cybersecurity incidents in this survey are as follows:

\begin{itemize}
\itemsep0em 
\item Private industry and flash notifications generated by the FBI~\cite{FBIa,FBIb,FBIc,FBId,FBIe}.
\item A list of significant cyber incidents provided by the Center for Strategic and International Studies (CSIS)~\cite{csisSignificantCyber} - a nonprofit policy research organization.
\item Internal reports or news updates from the affected organizations.
\item Available media sources that reported interviews with an official representative from the involved organizations.
\end{itemize}

The availability of the disclosed and documented security incidents is generally limited, that is due to their impact on the organization's reputation, finances, and services (i.e., in some cases, organizations prefer to not report these cases). Additionally, for some incidents, the FBI has not provided the names of organizations, hiding their identity. Considering the limited available information, we have summarised all the available incidents and provided details on financial loss, data loss, and service interruption to their consumers.

This survey is organized as follows: Section 2 describes 30 disclosed and documented cybersecurity incidents in the agriculture sector. Considering the available information, a brief description of the security threat, ransom amount, and impact on the organization are discussed in Section 2. Next, Section 3 reports popular cybersecurity frameworks and agriculture-specific cybersecurity solutions. These solutions could minimize the cyber risks and prevent or mitigate the cyberattacks. Section 4 explains the need to incorporate AI assurance and a novel AI assurance framework for the AI lifecycle of agricultural applications. Further, Section 5 discusses the effects of cybersecurity incidents in agriculture with relevance to critical infrastructures, food security, and the economy. The challenges to incorporate AI are also discussed in Section 5. Lastly, Section 6 provides an epilogue for this survey.

\section{Cybersecurity Incidents in Agriculture Sector}

This section details 30 malicious cybersecurity incidents disclosed and documented in agriculture. The summary of these incidents is provided in Table~\ref{tblr:incidence}, the details on each incident are as follows:

\begin{enumerate}

    \item \textbf{Monsanto, USA [July 2011, March 2014 \& January 2016]}
    
    Monsanto – an agrochemical and agricultural biotechnology corporation – reported three cyberattacks, resulting in data compromise. According to CNET~\cite{cnetMonsantoConfirms}, the first data breach happened in July 2011 that disrupted Monsanto's website. This data breach resulted in the data compromise of approximately 2500 individuals. Based on the Wall Street Journal~\cite{wsjMonsantoConfirms}, the second cyberattack occurred in March 2014 on one of the company's servers, exposing customer details such as credit card information and employee data. Additionally, this attack affected specialized seed-planting equipment, affecting about 1,300 farmers who were Monsanto's Precision Planting division customers. The third attack was reported in January 2016 on Climate Corp – a farm analytic firm acquired by Monsanto; resulting in compromising employee information and customers' credit card details. The Fox Business Network~\cite{foxbusinessMonsantoUnit} also noted that this affected more than 15000 farmers who own a Seed Sense 20/20 planting monitor.  

    \item \textbf{Target Corporation, USA [December 2013]}

    Target Corporation, the seventh-largest retail corporation in the U.S., confirmed a data breach on December 19, 2013. A statement released by Target Corporation~\cite{targetTargetConfirms} reported unauthorized access to payment card data in the U.S. stores and estimated impact on approximately 40 million payment card accounts between November 27, 2013, and December 15, 2013. Additionally, Target Corporation~\cite{targetUpdateData} reported that the data breach included customers’ details such as names, mailing addresses, phone numbers, or email addresses impacting up to 70 million individuals. A report submitted to the U.S. Securities and Exchange Commission (SEC)~\cite{secDocument} reports \$202 million spent by Target Corporation on legal fees and other costs, including a settlement amount of \$18.5 million.

    \item \textbf{Wendy’s, USA [February 2016]}

    Wendy’s – an international fast food restaurant chain – found unusual payment card activities in some restaurants in February 2016. Wendy’s~\cite{wendysPaymentCard} confirmed the evidence of a malware attack on point-of-sales systems in May 2016, which disabled all their franchise restaurants where the malware was discovered. Also, Wendy’s accepted that this malware attack targeted the payment card data, which includes customers’ details such as cardholder name, credit or debit card number, expiration date, verification, and service codes. Based on a news article~\cite{restaurantb}, Wendy’s agreed to pay \$50 million to settle claims, and they ended up paying \$27.5 million after utilizing insurance claims.

    \item \textbf{Chipotle Mexican Grill, USA [April 2017]}
    
    Chipotle Mexican Grill – an international fast casual restaurant chain - reported a malware attack on their point-of-sales systems on April 25, 2017. Based on Chipotle Mexican Grill’s statement~\cite{chipotleChipotleMexican}, the malware attack happened between March 24, 2017, and April 18, 2017, on point-of-sales systems, and the goal was to get access to payment card data from some Chipotle and Pizzeria Locale restaurants. Further, the company said, “The malware searched for track data (which sometimes has cardholder name in addition to card number, expiration date, and internal verification code) read from the magnetic stripe of a payment card”.

    \item \textbf{Mondelez International, USA [June 2017]}
    
    Mondelez International is one of the largest snack companies in the world, operating in 80 countries. It faced the NotPetya malware attack in June 2017 that locked up 1700 servers and 24,000 laptops~\cite{therecordMondelezZurich}. Additionally, due to this malware attack, their logistic software crashed, responsible for scheduling deliveries and tracking invoices. It took weeks for Mondelez International to recover, resulting in a loss of around \$180 million~\cite{crosignani2023pirates,CompaniesThought}. 
    
    \item \textbf{Whole Foods Market, USA [August 2017]}
    
    Whole Foods Market - a grocery chain – reported a credit card data breach impacting 117 venues in August 2017~\cite{cyberpolicyDataBreach}. Based on CNBC News~\cite{cnbcWholeFoods}, credit card information was mainly stolen from taprooms and full table-service restaurants. These restaurants utilize a different point-of-sale system than the central checkout systems at Whole Foods~\cite{globalnewsAmazonownedWhole}. The company did not disclose the details, including targeted locations and number of customers. 

    \item \textbf{Takeaway.com, Germany [March 2020]}
    
    Takeaway.com is a German food delivery service that delivers food from more than 15,000 restaurants in Germany. It suffered a DDOS attack on March 18, 2020, which caused delays and stopped order processing~\cite{cisomagAttackersLaunch}. This negatively impacted more than 15,000 restaurant owners and all the customers. The attackers demanded 2 Bitcoins, but the sources did not specify whether the company had agreed to pay the ransom demand~\cite{bleepingcomputerFoodDelivery}.
    
    \item \textbf{Agromart Group, Canada [May 2020]}
    
    The Agromart Group supplies fertilizer, crop protection, seed products, and other agriculture services across Eastern Canada. The Foreman \& Company~\cite{foremancompanyAgromartSollio} – a law firm that represented Agromart Group – specified that the Agromart Group suffered a Sodinokibi/Revil ransomware attack on or around May 27, 2020, extracting 22,328 files from their computers. The Agromart Group continued its business operations during this attack, operating manually. When the Agromart Group declined to pay the ransom, the attackers advertised this data breach and auctioned the data on the dark web to the highest bidder~\cite{farmtario}.

    \item \textbf{Harvest Sherwood Food Distributors, USA [May 2020]}
    
    Harvest Sherwood Food Distributors is based in San Diego, California, and is one of the largest independent meat and food industry distributors. The company faced a major Sodinokibi/Revil ransomware attack around May 3, 2020, that stole critical data from them. Based on DarkOwl~\cite{REvilHackers} – a leading provider of darknet data – this critical data includes roughly 2300 files containing details such as cash-flow analysis, distributor data, business insurance content, vendor information, and a dataset containing scanned images of driver’s licenses. The attackers also threatened the company to distribute the data and demanded \$7.5 million in ransom~\cite{REvilRansomware}.    

    \item \textbf{International Food and Agriculture Company, USA (unidentified) [November 2020]}

    A ransomware attack was conducted by OnePercent Group using a phishing email with a malicious zip file attachment. This attached file included a Microsoft Word or Excel document, which, after opening, dropped the IcedID banking trojan to affect the system. The attackers used IcedID to download and install Cobalt Strike on the affected systems, and then data exfiltration was done using rclone from the victim’s network. The FBI~\cite{FBIa} stated that the attackers maintained the network access for up to a month before deploying ransomware, downloading several terabytes of data and encrypting hundreds of folders. This resulted in a significant impact on the company’s administrative systems. The attackers asked for a \$40 million ransom, but the company did not pay it and successfully restored their systems from backups.
    
    \item \textbf{Americold Logistics, USA [November 2020 \& April 2023]}
    
    Americold Logistics is one of the largest temperature-controlled supply chain companies that offers cold storage facilities to food producers, retailers, and service providers. Based on a report published by the DHS~\cite{dhsPublicPrivateAnalytic}, Americold Logistics hit a ransomware attack on November 16, 2020, and to contain the intrusion, the company shut down its network. This severely impacted their phone system, email, inventory management, and order fulfillment~\cite{bleepingcomputerColdStorage}. The second cyberattack on Americold Logistics happened in April 2023. Based on a report published by the SEC~\cite{secArt20230426}, Americold Logistics received evidence of a cybersecurity incident on April 26, 2023. After this, the company immediately applied containment measures by taking operations offline to secure its system and reduce disruptions. Further, the company canceled all the inbound deliveries for a week and only allowed critical outbounds of the food products, which were close to expiring~\cite{thecybersecuritytimesAmericoldNetwork}.
    
    \item \textbf{Campari Group, Italy [December 2020]}
    
    Campari Group is one of the largest companies in the spirit industry, having a portfolio of more than 50 brands. On December 4, 2020, the company announced~\cite{campari} that they were a victim of a Ragnar Locker ransomware attack and the attackers had stolen two terabytes of data containing the personal information of more than 6,000 employees, accounting data, and some business contracts. Additionally, it also caused the shutdown of IT services and networks. Bitdefender~\cite{bitdefenderCampariStaggers} – a cybersecurity company – mentioned that the attackers demanded \$15 million worth of cryptocurrency, but the company did not make any statements about ransom.

    \item \textbf{Kroger, USA [January 2021]}
    
     Kroger - a supermarket giant – confirmed a data security incident in February 2021, which occurred on January 23, 2021. This attack affected Accellion, Inc., in which an unauthorized person accessed certain files affecting Accellion’s file transfer service~\cite{Kroger}. Kroger used the Accellion software to transfer files related to HR data, pharmacy, and clinic customer information. Kroger reported that it affected approximately 2\% of their customers, and based on news published by Reuters~\cite{reutersKrogerAgrees}, Kruger agreed to pay \$5 million to resolve the data breach claims. 

    \item \textbf{Agricultural Farm (unidentified), USA [January 2021]}

    A ransomware attack was conducted on a farm in the U.S. by gaining access to the farm’s internal network. The attackers gained administrative access through compromised admin credentials, which temporarily shut down their farming operations. This temporary loss caused a \$9 million loss. The FBI privacy industry notification~\cite{FBIa} and other sources~\cite{therecordFarmLoses,worldgrainCyberCriminals} did not provide additional details on this security incident.

    \item \textbf{Molson Coors Beverage Company, USA [March 2021]}
    
    The Molson Coors Beverage Company is a Chicago-based beer brewing company that suffered a ransomware attack in March 2021. Based on the company’s statement~\cite{molsoncoorsMolsonCoors}, it got hit by the cyberattack on March 11, 2021, and to stop the spread of malware, the company took its system offline. During this time, the system was not operational to the employees’ causing delays and disruptions in the brewery operations, production, and shipments in the United Kingdom, Canada, and the U.S. The company or FBI does not disclose additional details on the ransomware attack.

    \item \textbf{John Deere \& Co., USA [May 2021 \& August 2022]}
    
    Sick Codes – a researcher – reported bugs in John Deere's apps and website. Based on the details provided by Vice~\cite{viceBugsAllowed}, these bugs could expose customers' data, such as the vehicle or equipment owner's name, their physical address, the equipment's unique ID, and their Vehicle Identification Number or VIN. In August 2022, Sick Codes presented a new jailbreak for John Deere \& Co. at the DefCon security conference in Las Vegas. Based on the information provided by WIRED~\cite{wiredJailbreakJohn}, this jailbreak allows one to take control of multiple models using touchscreens, giving root access to the tractor's system. As The Verge~\cite{thevergeHackerShows} noted, ``this exploit could potentially help farmers bypass software blocks that prevent them from repairing the tractor themselves, " creating security implications for right-to-repair movement.

    \item \textbf{JBS Foods USA, USA [May 2021]}

    JBS Food USA is the second largest beef, pork, and poultry producer in the U.S., with over 66,000 employees. On May 30, 2021, JBS Food USA~\cite{jbsfoodsgroupCyberattackMedia} was hit by a Sodinokibi/REvil ransomware attack, which impacted many servers in the US and Australia. This caused the company to shut down nine US-based plants for several days, affecting meat supply in the U.S. and causing an increase in meat prices by 25\%. This also caused the company to lose less than one day of food production, and the company confirmed the payment equivalent to \$11 million in ransom~\cite{jbsfoodsgroupCyberattackMedia}.

    \item \textbf{McDonald’s Corporation, USA, South Korea \& Taiwan [June 2021]}
    
    McDonald’s Corporation suffered unauthorized activity on its internal security system on June 11, 2021, and it took a week for unauthorized access to cut off from its system. Based on a report published by the DHS~\cite{dhsPublicPrivateAnalytic}, this incident was a data breach that exposed private information of restaurants, employees, and their customers in the U.S., South Korea, and Taiwan. The Wall Street Journal~\cite{wsjNewsExclusive} reported that the data breach exposed business contact information of U.S. employees and franchisees as well as data on restaurants, including seating capacity and square footage of play areas.  

    \item \textbf{Baking Company (unidentified), USA [July 2021]}
    
    A baking company in the U.S. faced a cyberattack on July 2, 2021, due to a ransomware attack against Kaseya software. The Sodinokibi/REvil ransomware was deployed through Kaseya Virtual Server Administrator (VSA), so the company lost access to its server, files, and applications~\cite{FBIa}. This incident caused the bakery to be closed for approximately a week, resulting in delays in production and shipping and causing damage to the company’s reputation. The attack on Kaseya~\cite{kaseya} impacted approximately 1,500 downstream businesses, and on July 5, 2021, Kaseya software reported the development of a software patch.
    
    

    \item \textbf{Crystal Valley, USA [September 2021]}
    
    Crystal Valley is a farm supply and grain marketing cooperative in southern Minnesota and northern Iowa that faced a ransomware attack on September 19, 2021. Based on the official statement from Crystal Valley~\cite{crystalvalleyImportantNotice}, the ransomware attack infected the computer systems, resulting in interrupting the daily operations of the company, making it unable to accept credit card (Visa, Mastercard, and Discover) payments via their payment gateway~\cite{crystalvalleyCyberAttackUpdate}. Also, due to this attack, the company could not mix fertilizers or complete orders for livestock feed~\cite{reutersMinnesotaGrain}. 

    \item \textbf{NEW Cooperative Inc. and four other grain companies, USA [September 2021]}
    
     NEW Cooperative Inc. is an Iowa-based farm services business with 60 locations in Iowa and over 8,000 members. Based on the company’s statements published in The Washington Post~\cite{washingtonpostRussianHackers} and Reuters~\cite{reutersIowaFarm}, NEW Cooperative Inc.’s computer systems were offline on September 20, 2021, to contain a ransomware attack. Further, the company temporarily shut down its Midwest Agronomic Professional Services\textsuperscript{TM} (MAPS) – a soil map software - as a precaution. The Washington Post also mentioned that the “…hackers threatened to affect the software controlling 40 percent of the nation’s grain production, as well as the feed schedule of 11 million animals”. The BlackMatter ransomware-as-a-service (raas) tool was used for this attack, and attackers demanded \$5.9 million in ransom.
    
    Based on the notification published by the FBI~\cite{FBId}, four other grain companies were targeted by attackers in addition to Crystal Valley and NEW Cooperative Inc. incidences. All these attacks took place between 15 September and 6 October 2021. These attacks included ransomware variants such as Conti, BlackMatter, Suncrypt, Sodinokibi, and Blackbyte. The attack and ransom details of the other four grain companies are not disclosed.    

    \item \textbf{Schreiber Foods, USA [October 2021]}
    
    Schreiber Foods is a Wisconsin-based dairy company that produces cream cheese, natural cheese, processed cheese, beverages, and yogurt. Bloomberg~\cite{bloombergBloombergRobot} reported a cyberattack on Schreiber Foods plants and distribution centers in October 2021, which forced them to close milk plants for several days. The Wisconsin State Farmer~\cite{wisfarmerSchreiberFoods} reported that “…hackers demanding a rumored \$2.5 million ransom to unlock their computer system”, but the additional details are not mentioned anywhere. This cyberattack resulted in a shortage of cream cheese in the U.S. market, especially during the holiday baking season~\cite{cnnSurprisingReason}.

    \item \textbf{KP Snacks, UK [January 2022]}
    
     KP Snacks is a market leader in producing snacks, and it is present in over 30 countries worldwide. Based on the BBC news~\cite{bbcSnacksHack}, KP Snacks discovered a ransomware attack on January 28, 2022. The Conti ransomware group took responsibility for this attack and threatened to publish stolen data on February 6, 2022~\cite{zdnetSnacksWith}. This data included “credit card statements, birth certificates, spreadsheets with employee addresses and phone numbers, confidential agreements, and other sensitive documents”~\cite{threatpostSnacksLeft}. The darknet webpage also had a countdown time, which warned the company to publish more data unless a ransom was paid. This attack affected the company’s IT and communication system leading to supply issues for certain products such as Skips, Nik Naks, Hula Hopps, McCoy’s Crips, and KP Nuts which was expected until the end of March 2022~\cite{theguardianShortageNuts}.

    \item \textbf{Feed Milling Company (unidentified), USA [February 2022]}
    
    The FBI's Cyber Division~\cite{FBId} informed that a feed milling company that provides agricultural services reported two cybersecurity incidents in February 2022. In this incident, an unauthorized actor accessed the company's system, but the attempts were detected by the company and stopped before encryption occurred. The FBI or any other sources did not provide complete details about the ransom attack or disclose the company's details.

    \item \textbf{Multi-state Grain Company (unidentified), USA [March 2022]}
    
    A LockBit 2.0 ransomware attack happened in the multi-state grain company at the beginning of the spring planting season. The FBI~\cite{FBId} notified that a multi-state grain company suffered a LockBit 2.0 ransomware attack in March 2022. This company processes grains and provides farmers with seeds, fertilizers, and other logistic services. Before this attack, the FBI also provided indicators of compromise associated with LockBit 2.0~\cite{FBIe}.

    \item \textbf{Italy’s Ministry of Agriculture Website, Italy [May 2022]}
    
    The CSIS~\cite{csisSignificantCyber} noted the information about a DDoS attack on Italy’s Ministry of Agriculture website, which made the website unavailable. Reuters~\cite{reutersProRussianHackers} noted that the hacker group “Killnet” launched the attack around 2000 GMT, and the website was offline for the next few hours. Some other government websites were also unavailable because of the DDoS attacks, but there was no data breach~\cite{wsjRussianHackers}.

    \item \textbf{Apetito Group (Wiltshire Farm Foods and Meals on Wheels), UK [June 2022]}
    
    Apetito Group is a Germany-based frozen food supplier that offers free meals to hospitals, childcare facilities, social welfare homes, and charities. Based on their official statement~\cite{Apetito}, Apetito Group suffered a cyberattack over the weekend of June 25/26, 2022, which affected their IT systems and operating ability for the short term. This hindered their capacity to produce and deliver food for five days. The company's official statement did not provide details on this attack but ensured that the customers' payment card details were not compromised or stolen from this attack. This attack significantly affected their Wiltshire Farm Foods and Meals on Wheels operations, which are part of the Apetito Group.

    \item \textbf{Maple Leaf Foods, Canada [November 2022]}
    
    Maple Leaf Foods - Canada’s largest prepared meats and poultry producer - confirmed a system outage linked to a cybersecurity incident on November 6, 2022. The company took immediate action but stated that the “full resolution of the outage will take time and result in some operational and service disruptions”~\cite{mapleleaffoodsConfirmsSystem}. Maple Leaf Foods did not disclose any other details on this attack, but other sources~\cite{securityweekRansomwareGang,itworldcanadaMapleLeaf} considered the Black Basta ransomware group responsible for this incident. Further, WATTPoultry~\cite{wattagnetCyberattackCost} – a Canadian agri-food media – estimated a cost of at least CA\$23 million because of this attack on the company. This estimation was based on the financial results of the fourth quarter of the fiscal year of 2022, in which the company has indicated a net loss of CA\$ 41.5 million.

    \item \textbf{Dole plc, USA [February 2023]}
    
    Dole plc - a fruit and vegetable producer - reported a sophisticated ransomware attack in February in an annual report to the US SEC~\cite{secDole20221231}. Based on the report, Dole plc mentioned that the attack had a limited impact on their operations, so they shut down plants for a day and kept all the shipments on hold. Based on USA Today~\cite{usatodayDoleCyberattack}, this attack affected the supply of prepackaged salads, resulting in a shortage. Additionally, Dole plc’s first quarter of financial results~\cite{doleplc} noted that this attack was disruptive for their fresh vegetables and Chilean business, which cost them a total of \$10.5 million, from which \$4.8 million was related to continuing operations after the attack.  

    \item \textbf{Irrigation attack, Israel [April 2023]}
    
    The Jerusalem Post~\cite{jpostCyberAttack} reported a cyberattack on the Galil Sewage Corporation's water and wastewater treatment control systems in the Jordan Valley. These controllers were used for irrigating fields, which were dysfunctional for a day. The sources and other details of this cyberattack are unknown. However, The Council on Foreign Relations~\cite{cfrConnectDots} noted a similar attack in May 2020, conducted by Iranian hackers, that affected the water flow of two rural districts in Israel affecting agriculture.
    
\end{enumerate}

\begin{longtblr}[
theme = fancy,
  caption = {A summary of malicious cybersecurity incidents disclosed and documented in agriculture between July 2011 and April 2023. Most of the attacks were caused by Ransomware, and Target Corporation faced the highest financial loss (\$202 million) due to data breach.},
  label = {tblr:incidence},
  note{*}={FBI non-disclosed incident}
]{
  width = \linewidth,
  colspec = {Q[58]Q[237]Q[125]Q[137]Q[167]Q[80]Q[131]},
  row{1} = {c},
  cell{2}{1} = {c},
  cell{2}{2} = {c},
  cell{2}{3} = {c},
  cell{2}{4} = {c},
  cell{2}{5} = {c},
  cell{2}{6} = {c},
  cell{2}{7} = {c},
  cell{3}{1} = {c},
  cell{3}{2} = {c},
  cell{3}{3} = {c},
  cell{3}{4} = {c},
  cell{3}{5} = {c},
  cell{3}{6} = {c},
  cell{3}{7} = {c},
  cell{4}{1} = {c},
  cell{4}{2} = {c},
  cell{4}{3} = {c},
  cell{4}{4} = {c},
  cell{4}{5} = {c},
  cell{4}{6} = {c},
  cell{4}{7} = {c},
  cell{5}{1} = {c},
  cell{5}{2} = {c},
  cell{5}{3} = {c},
  cell{5}{4} = {c},
  cell{5}{5} = {c},
  cell{5}{6} = {c},
  cell{5}{7} = {c},
  cell{6}{1} = {c},
  cell{6}{2} = {c},
  cell{6}{3} = {c},
  cell{6}{4} = {c},
  cell{6}{5} = {c},
  cell{6}{6} = {c},
  cell{6}{7} = {c},
  cell{7}{1} = {c},
  cell{7}{2} = {c},
  cell{7}{3} = {c},
  cell{7}{4} = {c},
  cell{7}{5} = {c},
  cell{7}{6} = {c},
  cell{7}{7} = {c},
  cell{8}{1} = {c},
  cell{8}{2} = {c},
  cell{8}{3} = {c},
  cell{8}{4} = {c},
  cell{8}{5} = {c},
  cell{8}{6} = {c},
  cell{8}{7} = {c},
  cell{9}{1} = {c},
  cell{9}{2} = {c},
  cell{9}{3} = {c},
  cell{9}{4} = {c},
  cell{9}{5} = {c},
  cell{9}{6} = {c},
  cell{9}{7} = {c},
  cell{10}{1} = {c},
  cell{10}{2} = {c},
  cell{10}{3} = {c},
  cell{10}{4} = {c},
  cell{10}{5} = {c},
  cell{10}{6} = {c},
  cell{10}{7} = {c},
  cell{11}{1} = {c},
  cell{11}{2} = {c},
  cell{11}{3} = {c},
  cell{11}{4} = {c},
  cell{11}{5} = {c},
  cell{11}{6} = {c},
  cell{11}{7} = {c},
  cell{12}{1} = {c},
  cell{12}{2} = {c},
  cell{12}{3} = {c},
  cell{12}{4} = {c},
  cell{12}{5} = {c},
  cell{12}{6} = {c},
  cell{12}{7} = {c},
  cell{13}{1} = {c},
  cell{13}{2} = {c},
  cell{13}{3} = {c},
  cell{13}{4} = {c},
  cell{13}{5} = {c},
  cell{13}{6} = {c},
  cell{13}{7} = {c},
  cell{14}{1} = {c},
  cell{14}{2} = {c},
  cell{14}{3} = {c},
  cell{14}{4} = {c},
  cell{14}{5} = {c},
  cell{14}{6} = {c},
  cell{14}{7} = {c},
  cell{15}{1} = {c},
  cell{15}{2} = {c},
  cell{15}{3} = {c},
  cell{15}{4} = {c},
  cell{15}{5} = {c},
  cell{15}{6} = {c},
  cell{15}{7} = {c},
  cell{16}{1} = {c},
  cell{16}{2} = {c},
  cell{16}{3} = {c},
  cell{16}{4} = {c},
  cell{16}{5} = {c},
  cell{16}{6} = {c},
  cell{16}{7} = {c},
  cell{17}{1} = {c},
  cell{17}{2} = {c},
  cell{17}{3} = {c},
  cell{17}{4} = {c},
  cell{17}{5} = {c},
  cell{17}{6} = {c},
  cell{17}{7} = {c},
  cell{18}{1} = {c},
  cell{18}{2} = {c},
  cell{18}{3} = {c},
  cell{18}{4} = {c},
  cell{18}{5} = {c},
  cell{18}{6} = {c},
  cell{18}{7} = {c},
  cell{19}{1} = {c},
  cell{19}{2} = {c},
  cell{19}{3} = {c},
  cell{19}{4} = {c},
  cell{19}{5} = {c},
  cell{19}{6} = {c},
  cell{19}{7} = {c},
  cell{20}{1} = {c},
  cell{20}{2} = {c},
  cell{20}{3} = {c},
  cell{20}{4} = {c},
  cell{20}{5} = {c},
  cell{20}{6} = {c},
  cell{20}{7} = {c},
  cell{21}{1} = {c},
  cell{21}{2} = {c},
  cell{21}{3} = {c},
  cell{21}{4} = {c},
  cell{21}{5} = {c},
  cell{21}{6} = {c},
  cell{21}{7} = {c},
  cell{22}{1} = {c},
  cell{22}{2} = {c},
  cell{22}{3} = {c},
  cell{22}{4} = {c},
  cell{22}{5} = {c},
  cell{22}{6} = {c},
  cell{22}{7} = {c},
  cell{23}{1} = {c},
  cell{23}{2} = {c},
  cell{23}{3} = {c},
  cell{23}{4} = {c},
  cell{23}{5} = {c},
  cell{23}{6} = {c},
  cell{23}{7} = {c},
  cell{24}{1} = {c},
  cell{24}{2} = {c},
  cell{24}{3} = {c},
  cell{24}{4} = {c},
  cell{24}{5} = {c},
  cell{24}{6} = {c},
  cell{24}{7} = {c},
  cell{25}{1} = {c},
  cell{25}{2} = {c},
  cell{25}{3} = {c},
  cell{25}{4} = {c},
  cell{25}{5} = {c},
  cell{25}{6} = {c},
  cell{25}{7} = {c},
  cell{26}{1} = {c},
  cell{26}{2} = {c},
  cell{26}{3} = {c},
  cell{26}{4} = {c},
  cell{26}{5} = {c},
  cell{26}{6} = {c},
  cell{26}{7} = {c},
  cell{27}{1} = {c},
  cell{27}{2} = {c},
  cell{27}{3} = {c},
  cell{27}{4} = {c},
  cell{27}{5} = {c},
  cell{27}{6} = {c},
  cell{27}{7} = {c},
  cell{28}{1} = {c},
  cell{28}{2} = {c},
  cell{28}{3} = {c},
  cell{28}{4} = {c},
  cell{28}{5} = {c},
  cell{28}{6} = {c},
  cell{28}{7} = {c},
  cell{29}{1} = {c},
  cell{29}{2} = {c},
  cell{29}{3} = {c},
  cell{29}{4} = {c},
  cell{29}{5} = {c},
  cell{29}{6} = {c},
  cell{29}{7} = {c},
  cell{30}{1} = {c},
  cell{30}{2} = {c},
  cell{30}{3} = {c},
  cell{30}{4} = {c},
  cell{30}{5} = {c},
  cell{30}{6} = {c},
  cell{30}{7} = {c},
  cell{31}{1} = {c},
  cell{31}{2} = {c},
  cell{31}{3} = {c},
  cell{31}{4} = {c},
  cell{31}{5} = {c},
  cell{31}{6} = {c},
  cell{31}{7} = {c},
  hlines,
  vlines,
}
\textbf{Year} & \textbf{Incident} & \textbf{Country} & {\textbf{Attack }\\\textbf{ Cause}} & {\textbf{Financial }\\\textbf{Loss }\\\textbf{(Millions of~USD)}} & {\textbf{Data}\\\textbf{Loss}} & {\textbf{Service}\\\textbf{Interruption}}\\
{2011 \\ 2014\\ 2016} & Monsanto & USA & Data Breach & - & \ding{51} & \ding{55}\\
2013 & Target Corporation & USA & Data Breach & 202 & \ding{51} & \ding{55}\\
2016 & Wendy's & USA & Malware & 50 & \ding{51} & \ding{55}\\
2017 & Chipotle Mexican Grill & USA & Malware & - & \ding{51} & \ding{55}\\
2017 & Mondelez International & USA & Malware & 180 & \ding{51} & \ding{55}\\
2017 & Whole Foods Market & USA & Data Breach & - & \ding{51} & \ding{55}\\
2020 & Takeway.com & Germany & DDoS & - & \ding{55} & \ding{51}\\
2020 & Agromart Group & Canada & Ranomsware & - & \ding{51} & \ding{55}\\
2020 & {Harvest Sherwood Food Distributors} & USA & Ranomsware & 7.5 & \ding{51} & \ding{55}\\
2020 & {International Food and Agriculture Company*} & USA & Ranomsware & 40 & \ding{51} & \ding{55}\\
{2020 \\ 2023} & Americold Logistics & USA & Ranomsware & - & \ding{55} & \ding{51}\\
2020 & Campari Group & Italy & Ranomsware & 15 & \ding{51} & \ding{51}\\
2021 & Kroger & USA & Data Breach & 5 & \ding{51} & \ding{55}\\
2021 & Agricultural Farm* & USA & Ranomsware & 9 & \ding{55} & \ding{51}\\
2021 & {Molson Coors Beverage Company} & USA & Ranomsware & - & \ding{55} & \ding{51}\\
{2021 \\ 2022} & John Deere \& Co. & USA & Software Bugs & - & \ding{55} & \ding{55}\\
2021 & JBS Foods & USA & Ranomsware & 11 & \ding{55} & \ding{51}\\
2021 & McDonald’s Corporation & {USA, \\South Korea, \\Taiwan} & Data Breach & - & \ding{51} & \ding{55}\\
2021 & Baking Company* & USA & Ranomsware & - & \ding{55} & \ding{51}\\
2021 & Crystal Valley & USA & Ranomsware & - & \ding{55} & \ding{51}\\
2021 & NEW Cooperative Inc. & USA & Ranomsware & 5.9 & \ding{55} & \ding{51}\\
2021 & Schreiber Foods & USA & Ranomsware & 2.5 & \ding{55} & \ding{51}\\
2022 & KP Snacks & UK & Ranomsware & - & \ding{51} & \ding{51}\\
2022 & Feed Milling Company* & USA & Ranomsware & - & \ding{55} & \ding{55}\\
2022 & Multi-state Grain Company* & USA & Ranomsware & - & \ding{55} & \ding{55}\\
2022 & {Italy’s Ministry of Agriculture Website} & Italy & DDoS & - & \ding{55} & \ding{51}\\
2022 & Apetito Group & UK & Cyber Attack & - & \ding{55} & \ding{51}\\
2022 & Maple Leaf Foods & Canada & Ranomsware & 30.7 & \ding{55} & \ding{51}\\
2023 & Dole plc & USA & Ranomsware & 10.5 & \ding{55} & \ding{51}\\
2023 & Irrigation Attack & Israel & Cyber Attack & - & \ding{55} & \ding{51}
\end{longtblr}

\section{Cybersecurity Frameworks (CSFs)}

The effects of cyberattacks on organizations expose a need to implement a CSF that aims to minimize and prevent cyberattacks~\cite{hathaway2014best}. These frameworks are fundamental elements that include a collection of best practices that can guide farmers in identifying, detecting, and responding to cyberattacks~\cite{knapp2017maintaining,yigit2021cybersecurity}. This section discusses popular CSFs used in all sectors to minimize cyber risks. Additionally, recent state-of-the-art cybersecurity approaches specific to the agriculture sector are presented in sub-section 3.2.


\subsection{Generic CSFs}

A recent survey conducted by the Cloud Security Alliance~\cite{SecureDevOps} reported that the National Institute of Standards and Technology’s (NIST) CSF is the most adopted framework, while the Center of Internet Security’s (CIS) Foundation Benchmarks is the second most used. The International Organization for Standardization (ISO) and the International Electrotechnical Commission (IEC) published ISO/IEC 27000 security standards, which are also commonly used in different organizations~\cite{tofan2011information}. Recently, the U.S. Cybersecurity and Infrastructure Security Agency (CISA)~\cite{cisaBestPractices} recommended using the MITRE’s Adversarial Tactics, Techniques, \& Common Knowledge (ATT\&CK) CSF to develop adversary profiles, conduct activity trend analysis, and augment reports for detection, response, and mitigation purposes. ATT\&CK has also been used by the North Atlantic Treaty Organization (NATO) for Cyber Threat Intelligence (CTI) and modeling threats~\cite{parmar2019use,fox2018enhanced}. Considering these reasons, NIST, CIS, ISO 27000, and ATT\&CK are reported in this sub-section. A summary of these selected CSFs is presented in Table~\ref{tblr:gencsf}.

\subsubsection{NIST CSF 2.0}

The NIST CSF is a part of the Cybersecurity Enhancement Act, which is mandatory for federally identified or critical infrastructure sectors in the U.S.~\cite{force2013executive,fischer2014cybersecurity}. It describes essential cybersecurity outcomes that can be used to understand, assess, prioritize, and communicate cybersecurity risks in organizations. Based on the official documentation~\cite{pascoe2023public}, NIST CSF applies to all the information and communication technology infrastructures, information technology, IoT, and operation technology. However, it is not a one-size-fits-all framework to reduce cybersecurity risks. NIST also recommends using this framework with other resources to manage cybersecurity. The NIST CSF consists of four elements: Functions, Categories, Subcategories, and Informative References; and four implementation tiers: Partial, Risk Informed, Repeatable, and Adaptive. The five essential functions: Identify, Protect, Detect, Respond, and Recover, are aligned with 23 categories which are split between these functions. The details on the five key functions and categories are provided in Table~\ref{tblr:nsf}. The categories provided in Table~\ref{tblr:nsf} are designed to connect with 108 outcome-driven subcategories for enhancing the cybersecurity program.

\begin{longtblr}[
    theme = fancy,
  caption = {A brief description of the most popular CSFs used in organizations.},
  label = {tblr:gencsf}
]{
  width = \linewidth,
  colspec = {Q[165]Q[246]Q[527]},
  row{1} = {c},
  row{2} = {c},
  row{3} = {c},
  row{4} = {c},
  row{5} = {c},
  vlines,
  hline{1-2} = {1-3}{},
  hline{3-6} = {-}{},
}
\textbf{Framework} & \textbf{Developer} & \textbf{Description}\\
NIST 2.0 & U.S. Department of Commerce & {It includes essential cybersecurity outcomes that can be used to understand, assess, prioritize, and communicate cybersecurity risks in organizations~\cite{pascoe2023public}}\\
{Foundation\\ Benchmarks} & Center of Internet Security & {It includes recommendations to elevate the security of different cloud platforms, databases, containers, desktop software, mobile and network devices, and operating systems~\cite{cisecurityBenchmarksOverview}}\\
ISO/IEC 2700 Series & {International Organization for Standardization and International Electrotechnical Commission} & {It provides standards for Information Security Management Systems to become risk-aware for identifying and addressing weaknesses in organizations~\cite{disterer2013iso}}\\
ATT\&CK & MITRE & {It offers a knowledge base of tactics and techniques that represent the behaviors among the adversary lifecycle~\cite{strom2018mitre}}
\end{longtblr}

\begin{longtblr}[
  theme = fancy,  
  caption = {Functional goals and categories for each function are present in the NIST CSF. The goal descriptions and categories are taken from the NIST CSF documentation~\cite{pascoe2023public,framework2021getting}.},
  label = {tblr:nsf}
]{
  width = \linewidth,
  colspec = {Q[90]Q[445]Q[404]},
  row{1} = {c},
  row{2} = {c},
  row{3} = {c},
  row{4} = {c},
  row{5} = {c},
  row{6} = {c},
  vlines,
  hline{1-7} = {-}{},
  hline{2} = {2-3}{},
}
\textbf{Function} & \textbf{Goal} & \textbf{Category}\\
Identify & {Organizational understanding to manage cybersecurity risk to systems, assets, data, and capabilities.} & {Asset Management, Business Environment, Governance, Risk Assessment, Risk Management Strategy, Supply Chain Risk Management}\\
Protect & {Develop and implement the appropriate safeguards to ensure the delivery of services.} & {Identify Management \& Awareness, Data Security, Information Protection Processes \& Maintenance, Protective Technology}\\
Detect & {Develop and implement the appropriate activities to identify the occurrence of a cybersecurity event.} & {Anomalies \& Security Continuous Monitoring, Detection   Processes}\\
Respond & {Develop and implement the appropriate activities to act regarding a detected cybersecurity event.} & {Response   Planning, Communication, Analysis, Mitigation, Improvements}\\
Recover & {Develop and implement the appropriate activities to maintain plans for resilience and to restore any capabilities or services that were impaired due to a cyber-security event.} & {Recovery Planning, Improvements, Communications}
\end{longtblr}

\subsubsection{CIS Foundation Benchmarks}

CIS is a community-driven nonprofit organization whose mission is to develop, validate, and promote the best solutions to protect from cyber threats. CIS offers resources such as Controls, Benchmarks, SecureSuite, and Multi-State Information Sharing and Analysis Center (MS-ISAC) for safeguarding public and private organizations. CIS Benchmarks is the second most commonly used CSF, which provides recommendations to elevate the security of different cloud platforms, databases, containers, desktop software, mobile and network devices, and operating systems. Based on the official resource~\cite{cisecurityBenchmarksOverview}, the development of the benchmarks is a seven-step process: engage the community, define scope, create a draft, consensus begins, discuss and adjust, consensus met, and publish CIS Benchmark. This process involves subject matter experts, technology vendors, public, private and academic community members. Every recommendation in a benchmark provides a detailed discussion of profile applicability, description of the recommendation, rationale, impact, audit of the recommendation, and remediation. The organizations can select a suitable benchmark based on their technology, and additional CIS Foundation Benchmarks can be found on their official website.  

\subsubsection{ISO/IEC 27000 series}

ISO/IEC 27000 series is a family of standards for Information Security Management Systems (ISMS) using which organizations can become risk-aware to identify and address weaknesses~\cite{isoISOIEC27000}. Additionally, a certification against these standards can promote trust and significantly can help during legal disputes to reduce the risks of fines or compensations~\cite{pelnekar2011planning}. Disterer~\cite{disterer2013iso} reported that the ISO/IEC 27000 standards refer to the process-oriented Plan-Do-Check-Act (PDCA) cycle, which can be used to reduce risks and prevent security breaches. 
The ISO/IEC 27000 provides 39 control objectives and 134 measures to help with security management in organizations. 

\subsubsection{ATT\&CK CSF}

The ATT\&CK framework is a knowledge base of tactics and techniques that represents the behaviors among the adversary lifecycle, and these behaviors correspond to four granular levels: Tactics, Techniques, Sub-techniques, and Procedures~\cite{cisaBestPractices}. Based on the design and philosophy of ATT\&CK~\cite{strom2018mitre}, Tactics represent the “why” that focuses on the adversary’s technical goals, the reason for an action, and what they are trying to achieve. The Techniques answer “how” an action achieves a tactical goal, while Sub-techniques give a granular description of techniques. Lastly, Procedure indicates “what” an adversary did and how they used a Technique or Sub-technique. Overall, the ATT\&CK framework for enterprise includes 14 tactics, 193 techniques, and 401 sub-techniques. Companies such as Cisco~\cite{ciscoCiscoSecure}, Fortnite~\cite{Fortinet}, Claroty~\cite{clarotyClarotySupports}, and AttackIQ~\cite{attackiqLeveragingMITRE} have signified the importance of using ATT\&CK for CTI. Arista~\cite{Arista} recently documented which tactics and techniques from the ATT\&CK framework can be mapped to ransomware attacks. MITRE~\cite{mitre} also mapped REvil – a ransomware variant – with the ATT\&CK framework. The descriptions of tactics are provided in Table \ref{tblr:table1}; details on them can be found in Storm et al.~\cite{strom2018mitre,strom2017finding} and Pennington et al.~\cite{pennington2019getting}.

\subsection{Agriculture-specific Cybersecurity Solutions}

The advancement of technology allows farmers to collect and store large amounts of data about soil, crops, water, and other vital parameters required for managing agricultural systems~\cite{idoje2021survey}. Recent surveys~\cite{shaikh2022towards,liu2020industry,yang2021survey,akkem2023smart} on smart and precision farming have indicated the use of AI for different applications in the agriculture sector. For instance, the Convolutional Neural Network (CNN) and Artificial Neural Network (ANN) have been used for yield predictions~\cite{you2017deep}, weed detection~\cite{sa2017weednet}, and animal classification~\cite{hansen2018towards}. Additionally, advanced AI techniques have been applied in farm management applications such as predicting water and electricity consumption~\cite{shine2018machine} and optimizing water usage~\cite{dela2017water}. The AI has also been integrated with Instruction Detection Systems (IDSs) and other security solutions such as encryption, authentication, authorization, and blockchain to detect malicious behavior~\cite{ferrag2020deep}. These systems are needed to continuously monitor computer networks or system traffic to detect suspicious activities and provide alerts~\cite{ferrag2020deep}. Recently, the literature has pointed to a rise in the application of Deep Learning (DL)~\cite{ferrag2020deep,kwon2019survey,al2020survey,liu2019machine,ahmad2021network,wang2022deepfarm,gurrapu2021deepag} and Machine Learning (ML)~\cite{mohammadi2021comprehensive,buczak2015survey,mishra2018detailed} for detecting cyber security threats. Ferrag et al.~\cite{ferrag2020deep} surveyed different DL and ML techniques that detect intrusions in emerging technologies such as industrial agriculture, farming drones, and autonomous tractors. Considering these aspects, this section reports the extension of the survey conducted by Ferrag et al.~\cite{ferrag2020deep} by including recent cybersecurity solutions. These solutions can be integrated with CSFs for detecting intrusions. Table~\ref{tblr:agaisol} provides a summary of these cybersecurity solutions.

Two frameworks focused on intrusion detection for UAVs have been published in 2021 and 2023, respectively. Kumar et al.~\cite{kumar2021sp2f} proposed a Secured Privacy-Preserving Framework (SP2F) for UAVs, which are helpful for crop monitoring and capturing aerial images. In SP2F, the authors included a two-level privacy engine for data authentication and a stacked Long-Short Term Memory (LSTM) for anomaly detection. In the first level of the privacy engine, blockchain and smart contract-based enhanced Proof of Work (ePoW) are used for data authentication. In the second level, a Sparse Autoencoder (SAE) has been used to encode data to prevent inference attacks. Lastly, this data is given to a stacked LSTM for classifying attack and regular data instances. The authors used ToN-IoT~\cite{booij2021ton_iot} and IoT Botnet datasets~\cite{ullah2020technique} for experimentation, and performance was compared based on traditional classification metrics. The authors concluded that the SP2F framework outperformed the other related frameworks. The second framework is proposed by Fu et al.~\cite{fu2023machine} for UAV-assisted agricultural information security systems. The authors proposed a Double Deep Q-network (DDQN) algorithm based on Geography Position Information (GPI) for deploying UAVs. Authors argued that finding optimal deployment locations for UAVs is essential on farms because it can optimize the performance of wireless networks. Additionally, they combined Cuckoo Search (CS) with the Levy Flight to supplement UAV deployment. Levy Flight is a random walk-based sample generation technique used for UAV motion planning~\cite{shukla2022levy}. For an intrusion detection system, the authors integrated CNN and LSTM (CNN + LSTM), considering the data's time-series nature and solving the DL parameter explosion issue. The KDD-CUP99 dataset~\cite{tavallaee2009detailed} is used for experimentation, and the proposed method is evaluated using traditional classification metrics. The authors concluded that integrating CS and GPI balanced the performance while a CNN + LSTM architecture provided a 94.4\% intrusion detection rate.

\begin{longtblr}[
    theme = fancy,
  caption = {Descriptions of 14 tactics and the number of techniques present under them in the ATT\&CK framework. The descriptions presented in the table are based on Storm et al.~\cite{strom2018mitre} and Son et al.~\cite{son2023introduction}.},
  label = {tblr:table1}
]{
  width = \linewidth,
  colspec = {Q[167]Q[648]Q[125]},
  row{1} = {c},
  row{2} = {c},
  row{3} = {c},
  row{4} = {c},
  row{5} = {c},
  row{6} = {c},
  row{7} = {c},
  row{8} = {c},
  row{9} = {c},
  row{10} = {c},
  row{11} = {c},
  row{12} = {c},
  row{13} = {c},
  row{14} = {c},
  row{15} = {c},
  vlines,
  hline{1-16} = {-}{},
  hline{2} = {2-3}{},
}
\textbf{Tactic} & \textbf{Description} & {\textbf{Number of }\\\textbf{ Techniques}}\\
Reconnaissance & {Information gathering (active or passive) to plan an attack in the future that may include details of the victim organization, infrastructure, or staff/personnel} & 10\\
{Resource Development} & {Develop   resources involving creating, purchasing, or compromising/stealing resources that can be used to support attack} & 8\\
Initial Access & {Establish an initial connection required to attack targets in the network} & 9\\
Execution & Run a malicious code on a local or remote system & 14\\
Persistence & {Maintain access to the system across restarts, changed credentials, and other interruptions that could cut off access} & 19\\
{Privilege Escalation} & {Get higher-level permissions (root/local administrator) on a system or network} & 13\\
{Defense Evasion} & Avoid detection by defenders throughout the attack & 42\\
{Credential Access} & Obtain/steal account names, IDs, and passwords & 17\\
Discovery & Gain knowledge about internal networks and systems & 31\\
{Lateral Movement} & Enter and control a remote system on a network & 9\\
Collection & Collect data of interest to achieve Exfiltration & 17\\
{Command and\\Control} & Communication with the systems under the attacker’s control & 16\\
Exfiltration & Stealing data from the network & 9\\
Impact & Disrupt the availability or influence business operations & 13
\end{longtblr}

For network threat detection and classification, four recent frameworks are reported. In 2021, Peppes et al.~\cite{peppes2021performance} explored an ensemble method based on multi-model voting for detecting network threats. The authors included K-Nearest Neighbors (KNN), Support Vector Machine (SVM), Decision Tree (DT), Random Forest (RF), and Stochastic Gradient Descent (SGD) for this purpose. The predictions from these five models were used as inputs in hard and soft voting modules, which were then compared, and the accurate results were selected. The authors used the NSL-KDD dataset~\cite{tavallaee2009detailed} for experiments, and the attack detection accuracy was used for evaluation. Later, in 2022 two frameworks are proposed for network threat detection and classification. Smmarwar et al.~\cite{smmarwar2022deep} proposed a three-phase DMD-DWT-GAN malware detection framework for IoT-based smart farms. This framework utilizes two DL approaches: Generative Adversarial Network (GAN) and CNN. The first phase deals with the feature extraction and data preprocessing to obtain preprocessed data. This preprocessed data is then given to the second phase in which Discrete Wavelet Transform (DWT) is applied to generate Approximation ($A_c$), Detail ($D_c$), Vertical ($V_c$), and Horizontal ($H_c$) coefficients. The $D_c$ and $A_c$ coefficients are then fused with the generator and discriminator of the GAN model, respectively, to generate partial images. In the last phase, CNN uses these partial images and performs multi-class classification to identify the type of malware. The authors performed experiments on Malimg~\cite{nataraj2011malware} and IoT-malware~\cite{verma2020multiclass} datasets and reported an accuracy of 99.99\% on both datasets. Raghuvanshi et al.~\cite{raghuvanshi2022intrusion} used ML models for intrusion detection in smart crop irrigation, an essential component of Agriculture 4.0. The authors used Principal Component Analysis (PCA) for feature extraction and then compared the performance of the SVM, Logistic Regression, and RF techniques to detect intrusions. The model comparisons are based on traditional classification metrics, and the NSL-KDD dataset~\cite{tavallaee2009detailed} was used for the experiments. Based on the results, the authors concluded that SVM has the highest precision (98\%) compared to RF and logistic regression. In 2023, Javeed et al.~\cite{javeed2023intrusion} developed a novel DL-based IDS by combining Bidirectional  Gated Recurrent Unit (Bi-GRU) and LSTM with a softmax classifier to classify network attacks on smart farms. This approach is developed by considering the importance of edge computing in Agriculture 4.0 and detecting the attack at the network's edge. The authors also have used a Truncated Backpropagation Through Time (TBPTT) approach to handle long data sequences, which are common in network data. In the architecture, two Bi-GRU layers followed by two LSTM layers are used, which are then tested on three IoT datasets: CIC-IDS2018~\cite{sharafaldin2018toward}, ToN-IoT~\cite{booij2021ton_iot}, and Edge-IIoTset~\cite{ferrag2022edge}. The authors reported that the proposed IDS outperformed some baselines by achieving 99.82\%, 99.55\%, and 98.32\% accuracy for CIC-IDS2018, ToN-IoT, and Edge-IIoTset datasets. 

For DoS/DDoS attack detection in Agriculture 4.0, two frameworks have been reported. In 2021, Ferrang et al.~\cite{ferrag2021deep} developed DoS/DDoS attack detection frameworks to secure Agricultural 4.0 infrastructure. They used Recurrent Neural Network (RNN), CNN, and Deep Neural Network (DNN) – three different DL techniques – and compared their performance. For experiments, the authors used two datasets - CIC-DDoS2019~\cite{sharafaldin2019developing} and TON-IoT datasets~\cite{booij2021ton_iot} – and the performance of the models was compared based on traditional evaluation metrics such as accuracy, precision, recall, and ROC. The authors concluded that CNN provided an accuracy of 99.95\% for binary network traffic attack detection and 99.2\% for multiclass network traffic attack detection, outperforming the other two DL models. Later in 2023, Kethineni and Pradeepini~\cite{kethineni2023intrusion} developed a fused CNN model with a  Bi-GRU for detecting DoS/DDoS attacks. The data preprocessing - label encoding and data normalization - is performed in the input layer, which gives this preprocessed data to the Bi-GRU layer. The Bi-GRU layer consists of two GRUs and the attention layer. The attention layer is vital because it helps find the essential features to detect DDoS attacks. The output from the Bi-GRU layer is then given to the convolution layer to extract deeper features from the data. Lastly, the output layer, which consists of the SoftMax function, receives these features that classify the regular and DDoS attack instances. The authors tested this framework on APA-DDoS-Attack~\cite{sharafaldin2019developing} and TON-IoT datasets~\cite{booij2021ton_iot}, which resulted in 99.35\% and 99.71\% accuracy in attack detection.

In 2022, Privacy-encoding-based Federated Learning (PEFL)~\cite{kumar2021pefl}, and Federated Learning-based Intrusion Detection System (FELIDS)~\cite{friha2022felids} frameworks are proposed to ensure data security and privacy while performing network attack detection. The PEFL framework has two levels of data privacy and a Federated Learning integrated GRU (FedGRU) for intrusion detection. The first privacy level is based on perturbation-based encoding, responsible for feature mapping and normalization. Next, DL-based LSTM Autoencoder (LSTM-AE) is applied for data transformation in the second privacy level of the PEFL framework. Lastly, this transformed data is then given to the FedGRU module, responsible for classifying normal and attack data instances. The authors used the ToN-IoT dataset~\cite{booij2021ton_iot} for experimentation and compared performance using traditional classification metrics with other existing Federated Learning (FL) based techniques. Based on the results, the authors concluded that the PEFL framework improved the attack detection rate, precision, and F1 score for the ToN-IoT dataset compared to the non-FL and FL techniques. For the FELIDS framework, the motivation was to maintain data privacy via local learning, i.e., sharing the model updates with the aggregation server without sharing the entire data. In this framework, the authors used and compared three DL classifiers - DNN, CNN, and RNN. For experiments, the authors~\cite{friha2022felids} used three datasets - CSE-CIC-IDS2018~\cite{sharafaldin2018toward}, MQTTset~\cite{vaccari2020mqttset}, and InSDN~\cite{elsayed2020insdn} – and evaluated the performance using traditional classification metrics. They compared the performance of FELIDS with centralized DL models and noted that FELIDS provided the highest accuracy in detecting attacks.


\begin{longtblr}[
theme = fancy,
  caption = {Recent agriculture-specific cybersecurity solutions that combine different DL, ML, blockchain, and Federated Learning (FL) based techniques.},
  label = {tblr:agaisol}
]{
  width = \linewidth,
  colspec = {Q[46]Q[141]Q[146]Q[170]Q[150]Q[223]},
  row{1} = {c},
  row{2} = {c},
  row{3} = {c},
  row{4} = {c},
  row{5} = {c},
  row{6} = {c},
  row{7} = {c},
  row{8} = {c},
  row{9} = {c},
  row{10} = {c},
  row{11} = {c},
  vlines,
  hline{1-2} = {1-6}{},
  hline{3-12} = {-}{},
}
\textbf{Year} & \textbf{Authors} & \textbf{Framework} & \textbf{Techniques} & \textbf{Datasets} & \textbf{Objective}\\
2021 & {Kumar \\et al.~\cite{kumar2021sp2f}} & SP2F & {Blockchain, ePoW, LSTM, SAE} & {ToN-IoT\cite{booij2021ton_iot}, IoT Botnet~\cite{ullah2020technique}} & {Intrusion detection for UAV}\\
2021 & {Peppes \\et al.~\cite{peppes2021performance}} & - & {KNN, SVM, DT, RF, SGD} & NSL-KDD~\cite{tavallaee2009detailed} & {Network threat detection and classification}\\
2021 & {Ferrag\\et al.~\cite{ferrag2021deep}} & - & {RNN, CNN, DNN} & {CIC-DDoS2019~\cite{sharafaldin2019developing}, TONc~\cite{booij2021ton_iot}} & DoS/DDoS  attack detection\\
2022 & {Raghuvanshi\\et al.~\cite{raghuvanshi2022intrusion}} & - & {PCA, SVM, Logistic Regression, RF} & NSL KDD~\cite{tavallaee2009detailed} & {Network threat detection and classification}\\
2022 & {Smmarwar \\et al.~\cite{smmarwar2022deep}} & {DMD-DWT-GAN} & {DWT, GAN, CNN} & {Malimg~\cite{nataraj2011malware}, IoT-malware~\cite{verma2020multiclass}} & Malware detection\\
2022 & {Kumar\\et al.~\cite{kumar2021pefl}} & PEFL & {Perturbation-based encoding, LSTM-AE, FedGRU} & ToN-IoT~\cite{booij2021ton_iot} & {Data security,  privacy and network attack detection}\\
2022 & {Friha\\et al.~\cite{friha2022felids}} & FELIDS & {FL, DNN, CNN, RNN} & {IDS2018~\cite{sharafaldin2018toward}, MQTTset~\cite{vaccari2020mqttset}, InSDN~\cite{elsayed2020insdn}} & {Data privacy and network intrusion detection}\\
2023 & {Javeed\\et al.~\cite{javeed2023intrusion}} & - & {Bi-GRU, LSTM} & {CIC-IDS2018~\cite{sharafaldin2018toward}\\ToN-IoT~\cite{booij2021ton_iot}, Edge-IIoTset~\cite{ferrag2022edge}} & {Network threat detection and classification}\\
2023 & {Kethineni and \\Pradeepini~\cite{kethineni2023intrusion}} & - & {CNN, Bi-GRU, Wild Horse Optimization} & {ToN-IoT~\cite{booij2021ton_iot}, APA-DDoS~\cite{sharafaldin2019developing}} & DoS/DDoS attack detection\\
2023 & Fu et al.~\cite{fu2023machine} & - & {DDQN, GPI, CNN, LSTM, CS} & KDD-CUP99~\cite{tavallaee2009detailed} & Intrusion detection for UAV
\end{longtblr}


\section{Need of AI Assurance in Agriculture}

The use of AI has been observed significantly in different agriculture applications in addition to IDSs. Agriculture is a labor and time-intensive work evolving to increase efficiency and food production to satisfy the growing demand~\cite{saitone2017agri}. To achieve these objectives, many farms have adopted AI to perform agricultural automation~\cite{javaid2023understanding}. It has been noted that many farm machinery companies are using field equipment to capture data from the farms and transfer it to the centralized database~\cite{sudduth2020ai,gurrapu2021deepag,wang2022deepfarm,monken2021graph,batarseh2021public}. This data is then further used in AI models to assist farmers in decision-making. There are many agricultural AI applications such as production management, crop monitoring, crop disease detection, food quality assessment, intelligent spraying of pesticides, soil management, and harvesting~\cite{liu2020artificial,javaid2023understanding} where the collected has been used. Recently, Cross et al.~\cite{cross2018feed} developed an AI-based decision support tool that can detect changes in the feeding behavior of pigs and provide warnings of potential disease outbreaks. A decision-support tool like this can be used by farmers to take preventive actions that could increase the productivity of the animals. The other recent significant application of AI is irrigation management. King and Shellie~\cite{king2016evaluation} used ANN to predict the Crop Water Stress Index (CWSI). The objective was to develop a decision-support tool for wine grape production to assist with irrigation. The development of AI-driven decision-support tools has also been reported for crop nutrient management, especially to recommend optimum Nitrogen (N) fertilizer levels for corn~\cite{ransom2019statistical,qin2018application,tremblay2012corn,morris2018strengths}. The N fertilizers play a crucial role in corn production as the amount of N fertilizers affect the corn yield, affecting profit and may cause pollution~\cite{tremblay2012corn}. Ransom et al.~\cite{ransom2019statistical} compared eight different Machine Learning (ML) algorithms - Stepwise Regression, Ridge Regression, Least Absolute Shrinkage and Selection Operator, Elastic Net, Principal Component Regression, Partial Least Squares Regression, DT, and RF - to improve the existing N fertilizer recommendation tools. The authors used soil and weather data to develop the models and concluded that RF provided the best accuracy in all tested conditions. These examples indicate different applications of AI, which assist farmers in important farm-related aspects.

These decision-support tools can assist in providing operational support to farmers, but blindly following them could create problems. For example, 
let us assume that a farmer uses an AI-driven N fertilizer recommender tool. When the soil pH level – a soil characteristic parameter - is high, the tool recommends increasing the N input. This increases risks, and due to a lack of transparency, it is impossible to understand how the AI model is making those recommendations. Suppose an attacker performs a DoS/DDoS attack on the farm's network. In that case, the pH parameters can be compromised, causing data poisoning and affecting recommendations on the optimum amount of N fertilizer~\cite{ferrag2021deep}. This may result in massive losses to the farmer, as well as causing environmental pollution. Additionally, several AI challenges, such as model bias, adversarial attacks, data, and model security, affect process automation, resulting in reduced performance of IDSs for threat detection~\cite{laplante2020artificial}. This creates a need to validate and verify (i.e., assure) an AI tool that the users can trust. Further, agriculture is a primary pillar of the global bioeconomy~\cite{ams2023us}, which highlights the following: “AI must be ethical to make the right decisions, safe to protect users it potentially impacts, explainable so humans can understand it, fair in the decisions it makes, trustworthy so we have confidence in its abilities and secure to prevent cyberattacks and threats”~\cite{sobien2023ai}. Thus, it is crucial to deploy AI assurance as farms are getting more digitized to validate and verify the AI models used.

\begin{figure}[h]
\centering
\includegraphics[width=1\textwidth]{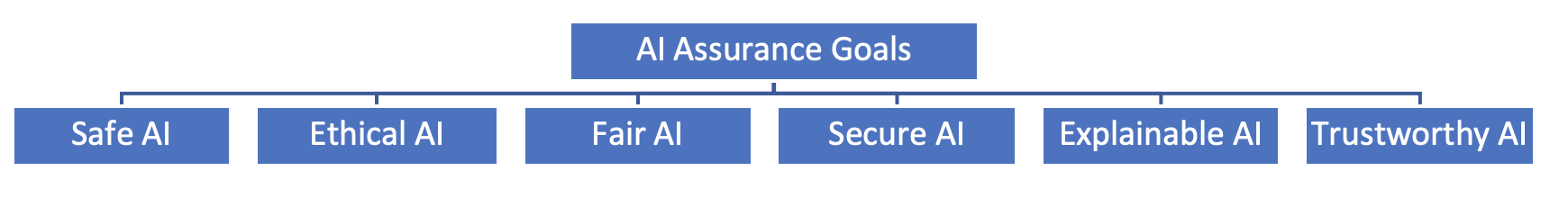}
\captionsetup{format=hang}
\caption{Six AI assurance goals that are needed for the verification and validation of AI.}
\label{fig:aigoal}
\end{figure}

AI assurance can be defined as a “process that is applied at all stages of the AI engineering lifecycle ensuring that any intelligent system is producing outcomes that are valid, verified, data-driven, trustworthy and explainable to a layman, ethical in the context of its deployment, unbiased in its learning, and fair to its users”~\cite{batarseh2021survey}. Based on Batarseh et al.~\cite{batarseh2021survey}, six goals of AI assurance include: Ethical AI, Safe AI, Fair AI, Secure AI, Explainable AI, and Trustworthy AI, which are shown in Figure~\ref{fig:aigoal}. These goals are needed for the assurance of AI. In October 2023, the president of the U.S. signed an executive order to promote Safe, Secure, and Trustworthy AI that aims to protect Americans from the potential risks of AI systems~\cite{whitehouseFACTSHEET}. Recently, the U.S. National Science and Technology Council (NSTC) considered AI as a national priority and prepared the National AI Research \& Development (R\&D) strategic plan~\cite{nitrdNATIONALARTIFICIAL} to manage the risks of AI. This plan includes seven strategies, four focusing on responsible AI, human-AI collaborations, ethical AI, and safe \& secure AI. The DHS~\cite{cisaSoftwareMust} also stated that AI is software, and so it must be secure by design. Further, the Centre for Data Ethics and Innovation (CDEI) of the UK government has developed a roadmap for AI and highlighted the need for AI assurance to manage AI risks~\cite {undefinedRoadmapEffective}. This roadmap provides the vision, challenges, and changes needed to develop an effective and mature AI assurance ecosystem. The research community has also observed similar interest, focusing on developing explainable, interpretable, and trustworthy AI~\cite{freeman2021enabling}~\cite{ali2023explainable}. These reasons make it essential to think of AI assurance as a part of AI development.

\subsection{The Farmer-Centered AI (FCAI) Framework}

One approach to achieving AI assurance goals is integrating human interaction in the AI lifecycle, known as Human-centered AI (HCAI), we restructure that and propose a Farmer-Centered AI, or the FCAI framework. In FCAI, farmers and AI get equal importance compared to the ``traditional AI lifecycle". FCAI focuses on farmers' inputs, interactions, and collaborations while designing, training, and evaluating AI~\cite{sperrle2021survey,spinner2019explainer}. For instance, UAVs are used in agriculture for different applications such as crop disease detection, making harvest-related decisions, surveillance, and predicting soil parameters~\cite{norasma2019unmanned,su2023ai}; but these can bring different challenges. Some of these challenges can arise regarding AI algorithms and farmers' acceptance. AI challenges, such as algorithm generalizability and labeled data scarcity depend on the AI developer~\cite{su2023ai}, while the farmers' acceptance solely depends on them. Jarman et al.~\cite{jarman2016unmanned} reported that the results generated from the UAVs should provide actionable information for agricultural practice, and it is critical to translate these into actionable steps in the agriculture lifecycle. This example shows a need for engagement between AI developers and farmers to use AI appropriately. Thus, the involvement of farmers in this process can empower and enable them to foster ethical, interactive, and contestable use of AI while uncovering underlying biases and limitations~\cite{capel2023human}. Accordingly, FCAI framework can provide a high level of control to farmers while achieving automation of the agricultural lifecycle. As noted by Shneiderman~\cite{shneiderman2020human}, this approach can make AI reliable, safe, and trustworthy for decision-making, benefiting individuals, families, communities, businesses, and society. In our case, FCAI allows the understanding of farmers’ needs along with helping farmers to understand AI~\cite{riedl2019human}. This approach can assure transparency and accountability that will enhance public confidence, increasing societal, and commercial values~\cite{shneiderman2021human}. Considering these aspects, the FCAI framework is proposed and shown in Figure~\ref{fig:aiag}. This framework includes six AI assurance goals and five essential AI lifecycle components.


Smart farms use different IoT devices to collect data about soil, crops, irrigation, farm animals, weather, and other related variables. It has been noted that security is a foundational component required to protect the infrastructure from cyber security threats. Similarly, \textit{Secure AI} aims to prevent or mitigate cyber threats that can affect normal operations~\cite{sobien2023ai}. These reasons make it necessary to secure all framework components and, thus, are enclosed under the \textit{Secure AI} goal. Essentially, \textit{Secure AI} deals with intrusion detection, infrastructure protection, and threat mitigation. This also includes securing different network resources, such as UAVs, sensors, routers, and decision-making tools. Another important consideration in agricultural security is the rise of vertical and urban farms. After ensuring secure data collection, the data is provided to the data processing and assurance component. This is an essential step to clean and organize the data, which is needed to understand the biases, skewness, variances, and incompleteness in the collected data. It is also a critical step in anomaly detection and classification applications such as crop disease and weed detection, which involves imbalanced data that negatively affect the AI model~\cite{kulkarni2020foundations}. 

Additionally, \textit{Ethical AI} is injected into data processing and assurance to make AI recommend appropriate decisions to benefit people~\cite{ayling2022putting}. The ethical component can include elements such as the USDA farm bill and regulations on using fertilizers and pesticides. Further, the hardware components, such as sensors or drones, can cause harm to cattle, which makes it essential to incorporate animal welfare aspects and environmental protection policy data~\cite{mark2019ethics}. The other important consideration is the size of the farm; about 80\% of the U.S. farms are small farms~\cite{usdaUSDAFood}, and the decisions on crop selection, growth, and harvest can influence the supply of agricultural commodities, reflecting changes in economic markets and trade.~\cite{batarseh2021public, monken2021graph}. This makes it essential to include those relevant datasets regarding small and midsize producers offered by USDA. This external knowledge can help incorporate societal and commercial values along with tackling data biases. These elements are then combined and inputted to the Model Development component. The model development component focuses on developing ML and DL models that aim to generate predictions for assisting farmers in decision-making. The predictions generated from the ML and DL models are then evaluated. These evaluations are critical and considered as a part of algorithmic evaluation. 

For the evaluation, metrics, such as root mean squared error, and R$^2$, are used for regression, while precision, recall, and accuracy are used in classification tasks. These evaluations are needed to understand the performance of AI against the available data, which is then used for model improvement by performing model tuning. After selecting an appropriate AI model, the model's predictions are evaluated to understand fairness, trustworthiness, and explainability. The \textit{Fair AI} and \textit{Trustworthy AI} can be considered a part of algorithmic assurance because farmers evaluate the predictions. These predictions may include recommendations on crop rotations and harvesting affecting retail and food distribution. Thus, it is crucial to include farmers in this process to understand their needs and help them understand the use of AI. This involvement could bring confidence and trust in AI, which may result in Fairness. The \textit{Explainable AI} is achieved by explaining the predictions to the farmers and evaluating the explanations to understand farmers' perceptions. For this purpose, different local and global explanation methods can be used to improve understanding. It has been noted that \textit{Explainable AI} can help make AI a white-box model that increases trust, fairness, and confidence in using AI~\cite{adadi2018peeking}. This is vital for farmers, especially considering the use of AI in the bioeconomy sector for decision-making. Next, after fulfilling these requirements, the model is deployed, and access to the decision-support tool is provided to the farmers. Lastly, \textit{Safe AI} can be integrated at the last stage in the AI development, i.e., while using the decision-support tool. The \textit{Safe AI} aims to ensure the well-being of the users and those impacted by it~\cite{sobien2023ai}. Morgan~\cite{morgan2006decision} noted that the goal of the agriculture decision-support tool is to integrate social, economic, environmental, and cultural well-being while making decisions. Thus, by injecting the rules, data, and relevant information on sustainable resource management, the recommendations generated by
AI can be evaluated to ensure the life and well-being of farms, farmers, society, and environment. This way, the FCAI framework incorporates six goals of AI assurance into the AI development process for agriculture. This approach of integrating AI assurance with farming practices may benefit farmers in taking appropriate actions for farm management. Further, the predictions/recommendations from the decision-support tool can be utilized in crop water management, such as irrigation management, to supply the optimum quantity of water to crops at critical stages in the growing season~\cite{sudduth2020ai}. This entire process may affect the yield and further can affect the agribusiness.

\section{Discussion: Impacts of Cybersecurity Incidents and Challenges}

This section details the impact of cybersecurity incidents in agriculture on other critical infrastructures, food security, and the economy. Further, this section also explains the practical challenges affecting the utilization of AI in agriculture.

\subsection{Critical Infrastructure Sectors}

The agricultural sector is a node within the larger system of critical infrastructure sectors in a nation. This is because networks of supply chains and Water \& Wastewater Systems (WWS) connect with it. The previous sections reported vulnerabilities and impact on smart farms, but it has cascading effects that disrupt delivery to the Food Supply Chain (FSC)~\cite{demestichas2020survey}. A FSC includes a set of activities, as noted by Nukala et al.~\cite{nukala2016internet}, a ``farm-to-fork" sequence. This sequence involves farmers, food producers/processors, food quality assessment, logistics, packaging, warehousing, and marketing. Thus, it can be seen that the FCS depends on the smart farm to yield the crops and animal products that can be processed and distributed to ensure food security. Considering the perishable nature of most farm products (before processing) and specific harvest times in the year, there are a lot of uncertainties in the operation of the system. This includes the early links that form the dependencies on the entire system~\cite{barreto2018smart}. To help deal with these uncertainties, the FSCs are also adopting emerging technologies like IoT and Information and Communication Technologies (ICT), making it vulnerable to cyberattacks~\cite{alsinglawi2022internet, latino2022cybersecurity}. Additionally, attackers can attack the supply chain or third-party partners that provide raw materials such as fertilizers, grains, and stock feed affecting the farms~\cite{yazdinejad2021review}. This is the typical technological trade-off in critical infrastructure: benefits for better monitoring, logistics, and decision-making at the cost of opening to data breaches and ransomware. These technologies, however, are adopted throughout several stages of the supply chain. For example, Latino and Menegoli~\cite{latino2022cybersecurity} explain that the food and beverage industry is applying new technology to the stages of ``farming, transportation, processing and the packaging and storage of food products". In this, each layer has its technological network of interdependencies between cyber and physical components, and then strung together, forming a chain of dependencies. This opens up the supply chain to vulnerabilities through cascading effects of cyberattacks at any one stage and potentially impacting later stages that are secure. For example, the ransomware attacks noted on food and agriculture companies~\cite{foremancompanyAgromartSollio,FBIa,jbsfoodsgroupCyberattackMedia,crystalvalleyImportantNotice,reutersIowaFarm,FBId} in Section 2 are real examples of this situation. Additionally, the ransomware attacks on grain companies such as Crystal Valley~\cite{crystalvalleyImportantNotice} and NEW Cooperative Inc.~\cite{reutersIowaFarm} interrupted the daily operations of these companies. This would have affected the farmers because Minnesota and Iowa are the third and fourth largest producers of crops in the U.S., respectively.  

Another critical infrastructure sector that impacts agriculture and FSCs is the WWS sector. The WWS sector is one of the defining areas of critical infrastructure for agriculture because it provides water for irrigation, crop cooling, frost control, and pesticide \& fertilizer applications~\cite{ward2002economic}. The agricultural water makes it possible to grow fresh produce and sustain livestock. It has been noted that 37\% of the water goes towards irrigation of crops, which is the second largest use of water after thermoelectric power production~\cite{bryant2022water}. Recently, a technological shift has been observed toward adopting Industrial Control Systems (ICS) and Systems Control and Data Acquisition (SCADA) in water systems. These technologies, without appropriate security measures, make the water systems vulnerable to cyber threats~\cite {bryant2022water, hassanzadeh2020review, kayan2022cybersecurity}. Considering more than a third of our water supply is utilized for irrigation, any disruptions in WWS cascade the impact on agriculture, affecting food security and supply chains. Additionally, a subtle change in the chemical agents used in the WWS, for example, sodium hydroxide, can change the pH level of the water~\cite{grubbs2021evolution}. This affects water quality, crop yield, and food production. Recently, in Florida, an adversary tried to poison the water by changing the values of sodium hydroxide in the system~\cite{hassanzadeh2020review}. This happened due to the compromising of credentials. Further, incidents such as the irrigation attack~\cite{jpostCyberAttack}, Maroochy Shire Sewage Water Spill, where one million liters of sewage spilled in the local water system, and the Riviera Beach ransomware attack that disrupted the service of the water utilities~\cite{kayan2022cybersecurity} can create devastating impacts on agriculture and FSCs.

Lastly, as noted in Section 2, in May 2022, there was a DDoS attack on Italy's Minister of Agriculture website~\cite{csisSignificantCyber}. This attack made some government websites unavailable for a few hours. As noted by Geiller and Lee~\cite{lee2019using}, government websites are crucial for efficient public administration, effective public services, and maintaining the democratic legitimacy of the government. The government's agricultural website can provide different tools for farmers, which becomes unavailable due to cyberattacks on the government websites. For example, farmers.gov\footnote{https://www.farmers.gov/} offers local dashboards that provide insightful information on farming data and USDA resources for different U.S. states. Additionally, it offers different tools such as farm loan assistant tools, disaster assistance discovery tools, and conservation concern tools, which can help farmers in decision-making. Thus, a cyberattack on the government infrastructure, such as websites, could create delays in resource sharing and serving the country's agricultural community. 




\subsection{Food Security, Accessibility \& Availability}

Cybersecurity incidents in the agricultural sector profoundly impact food security. This disrupts various availability aspects, such as farm production, food processing, and supply chains. The complex and highly diverse enterprises within the food and agricultural system exacerbate food security vulnerability. Furthermore, the 

\begin{landscape}\centering
\vspace*{\fill}
\begin{figure}[h]
\centering
\includegraphics[width=1.6\textwidth]{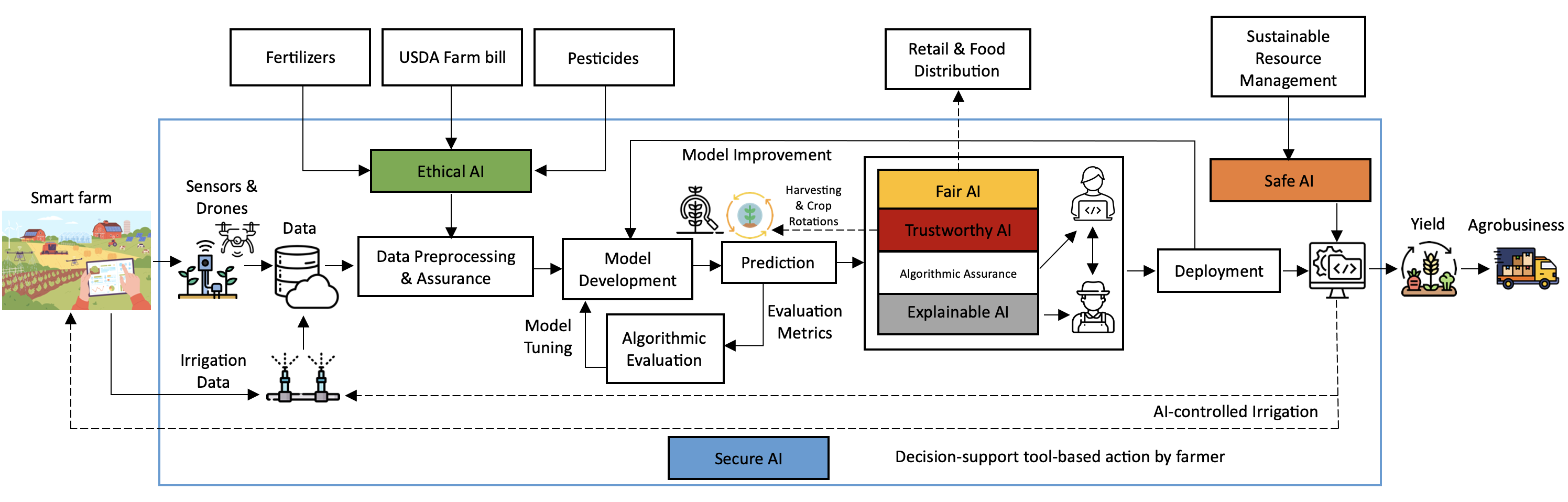}
\captionsetup{format=hang}
\caption{The FCAI framework incorporates six goals of AI assurance at different stages of the AI lifecycle injected into the agricultural lifecycle.}
\label{fig:aiag}
\end{figure}
\vfill
\end{landscape}

\noindent rapid integration of emerging technologies and their widespread adoption in the agriculture sector 
amplifies susceptibility to cybersecurity threats and potential cyberattacks~\cite{duncan2019cyberbiosecurity}.
The incidents reported in this survey illuminate the multifaceted challenges in the agricultural sector to maintain operational integrity. Farm production, food processing, and supply chains are vital to the global food security network, but their connections with emerging technologies make them vulnerable to cyber threats. This highlights the need for heightened vigilance and robust protective measures. The analyzed cyber incidents in the agricultural sector manifest a diverse array of attack vectors, ranging from ransomware attacks \cite{dhsPublicPrivateAnalytic, FBIa, jbsfoodsgroupCyberattackMedia, reutersMinnesotaGrain, Apetito} to unauthorized access through compromised credentials \cite{Kroger, wsjNewsExclusive} and sophisticated hacking techniques \cite{cisomagAttackersLaunch, csisSignificantCyber}. This diversity underscores the need for multifaceted security strategies to address a wide spectrum of cyber threats. Notably, these incidents resulted in substantial disruptions at various nodes within the agricultural supply chain, encompassing farm operations \cite{FBIa, jbsfoodsgroupCyberattackMedia, bloombergBloombergRobot}, processing facilities \cite{reutersMinnesotaGrain, secDole20221231}, and distribution \cite{cisomagAttackersLaunch, dhsPublicPrivateAnalytic} networks. It has also been noted that these attacks disrupted the communication networks, inventory management, and order fulfillment. This forced companies to apply containment measures, shut down operations, cancel or delay deliveries, and prioritize outbound shipments for expiring food products.

In the aftermath of cyberattacks on the agricultural sector, profound consequences impact essential food availability and accessibility. The cyberattacks targeting critical nodes within the sector, such as meat, dairy, frozen food, and restaurant supply chains, disrupt the flow of essential food items to consumers. These disruptions could occur throughout the entire production and distribution network. It causes shortages in crucial food categories compromising the variety and availability of products on store shelves \cite{bloombergBloombergRobot, secDole20221231}. Furthermore, farm supply shortages, including vital resources like grain and fertilizers, can occur due to cyberattacks directly affecting food production \cite{crystalvalleyImportantNotice, FBId}. 

In 2011, the FDA took decisive steps to implement the Food Safety Modernization Act (FSMA). The FSMA underlines the importance of safeguarding our food supply chain and provides rules to ensure their safety. These rules are not merely a regulatory obligation but a shared responsibility that spans numerous stages in the global supply chain. This recognition underscores the critical significance of food security in the broader context of international trade and public health \cite{FSMA_2023}. The FSMA emphasizes the need for comprehensive measures to guarantee the integrity and safety of the food supply, particularly in the face of emerging cybersecurity vulnerabilities. Recently, the U.S. government has increasingly recognized the intersection between national food security and cybersecurity. This awareness has prompted the formulation of specific policies and regulations to mitigate the vulnerabilities posed by cyber threats to the food supply chain. For instance, the FSMA in the U.S., enforced by the FDA, acknowledges the growing cyber threats. This understanding was further solidified by initiatives such as the Cybersecurity Assessment and Risk Management Approach (CARMA). The CARMA is a risk assessment program led by the DHS, FDA, and USDA, collaborating with industry, academia, and government representatives. These initiatives were influenced by insights from the Food Industry Cybersecurity Summit \cite{streng2016food} organized by the Food Protection and Defense Institute (FPDI) and supported by the DHS Science and Technology Directorate. In 2016, this summit was held in Washington, D.C., and facilitated discussions among nearly 40 participants from the food industry, government, and academia. These conversations shed light on critical cyber-related risks faced by the food industry, particularly concerning the extreme vulnerability of industrial control systems (ICS). Additionally,  inherent vulnerabilities within ICS, lack of awareness about the interaction between ICS and information technology (IT) systems, and dependence on outsourced technology management have been highlighted as significant concerns. 

To resolve those vulnerabilities, FPDI has outlined comprehensive short and long-term actions for industry stakeholders. Among these, fostering communication between Operations Technology (OT) and Information Technology (IT) staff is paramount. It further bridges the cultural gap and encourages a deeper understanding of ICS and IT system interactions. Additionally, the risk assessments encompassing ICS and IT systems are crucial, involving cybersecurity experts in procuring ICS devices. Furthermore, extending the robust food safety and food defense culture to cybersecurity initiatives is emphasized, recognizing the inherent link between security and safety. To achieve this collective progress, it is paramount to have industry-government partnerships in cyber-related standards organizations.

\subsection{Economy}

Cybersecurity has multiple components  - public good, externalities, informational asymmetries, and property rights - which challenge a formal cost-benefit analysis~\cite{bauer2016handbook}.  The summarised incidents reported in the survey show that affected parties tend to quantify immediate losses, but they often ignore other dimensions of the cost-benefit calculations. Investment in public goods or services with externalities tends to deviate from societal optimum, i.e., market failure, where investing companies create a wall to protect information and retain property rights. A complete enumeration of costs and benefits is difficult, but Lis and Mendel~\cite{lis2019cyberattacks} provided a long list of direct and indirect costs of cybersecurity attacks. The direct costs include - operations and continuity; income flows; physical security; lost customers, reputation and relations; equity market effects; legal issues, including service agreement lapses, insurance, IP losses; and others. In the compressed timeline addressing recovery and restoration, indirect costs are - lost productivity, future revenues, and shying away from future opportunities due to high risk - are often paid little attention.  The agriculture sector is unique and interesting since it already faces weather/climate and price risks.  That is, Cybersecurity is a third dimension to risks in the FA sector. 

The review of cybersecurity incidents in the paper indicates direct costs measured by these companies range from a few to \$202 million.  Much of it is income losses and disruptions to processes, with many other direct costs unaccounted for, e.g., legal issues arising from privacy loss.  Thinking along the lines of weather/climate and price risks, the cybersecurity attacks will have a comparable effect.  That is, if the risk is not addressed, attacks will result in higher prices along the entire supply chain affecting consumers, producers (farmers), processors, wholesalers, and retailers.  For example, Venkat and Masters~\cite{venkat2022retail} estimate price increases of up to 42 percent for fruits, vegetables, dairy, and eggs following storms and droughts.  
        
With squeezed profitability, since the agriculture sector does not have high pricing power (small and medium enterprises), business survival and job stability are the next set of economic challenges. Following up on the storms and drought example, the high prices hurt consumers and do not necessarily benefit farmers and others along the supply chain. The latter arises when prices of intermediates, e.g., energy and transportation costs, capture most of the margin between markets and production points at the time of weather events.  Additionally, reputational damage and stock market losses can make companies leave the agriculture sector disrupting domestic and global supply chains. 
        
With cybersecurity risks, farmers and others along the supply chain face significant uncertainties in the business environment.  On one side, they ask whether or not investments in cybersecurity are worthwhile given their scale, limited pricing power and the public good properties of such systems. Simultaneously, they recognize the consequences of an attack (direct and indirect costs). Uncertainties in business environment have been shown to affect investment, consumption, GDP and even trade among nations~\cite{khadka2023anomalies}.  A cybersecurity risk management strategy is the need of hour in the agriculture sector, and it will require a public and private partnership. 

\subsection{Challenges}

This sub-section discusses the challenges and potential solutions to improve the current state of utilization of AI in agriculture. Considering this, the three most relevant challenges are: Data, Deployment, and Adoption. The details of these challenges are as follows. 

\begin{itemize}
\itemsep0em 
    \item \textbf{Data} - Data plays a crucial role in developing AI solutions, and it is a significant bottleneck in getting labeled high-quality data to train AI models~\cite{roh2019survey}. Recent surveys published by Ferrang et al.~\cite{ferrag2021cyber} and Alwis et al.~\cite{de2022survey} listed different datasets that can be used for different smart farming applications. However, they are not specially obtained from agriculture or smart farms. It is essential to represent the characteristics of agriculture in the data to develop a robust AI solution. This creates a need to obtain high-quality domain-specific datasets. One of the possible solutions to make diverse, high-quality data available is using agriculture testbeds. Recently, Batarseh et al.~\cite{batarseh2023acwa} and Oberascher et al.~\cite{oberascher2022smart} reported soil testbeds that can help to produce agriculture-specific network and sensor data. This approach may help researchers to tackle data quality issues and provide robust AI solutions for agriculture.

    \item \textbf{Deployment} – Deployment of AI solutions in real settings is another challenge observed in the agriculture domain. Most security solutions, such as malware, DoS/DDoS detection systems, and intrusion detection for UAVs, are not deployed. These security solutions are crucial for detecting and mitigating cybersecurity threats to farms. The deployment of AI solutions can also uncover the scalability and stability issues that may occur while using these solutions. Additionally, issues such as data and model drifts can occur with the deployment of models, which also needs attention to use AI as a decision-support tool.

    \item \textbf{Adoption} - The adoption of AI in everyday decision-making for farmers is challenges because it depends on different factors such as personal, social, cultural, and economic~\cite{pannell2006understanding,howley2012factors}. Prokopy et al.~\cite{prokopy2008determinants} noted that the education level of farmers, capital, income, farm size, access to information, and environmental awareness could also affect adopting new technology. In addition, the missing link between AI and AI assurance could increase the risk of using AI at the farm, affecting farmers' trust, confidence, and safety. Thus, it is essential to incorporate AI assurance along with the deployment of AI applications and involve farmers in essential steps, as shown in FCAI framework.  Further, providing informative sessions on explaining the need for AI, its use, and real-case scenarios could also help increase AI adoption in agriculture.
    
\end{itemize}

\section{Epilogue}

The 30 FA incidents reviewed in this survey spanned over 11 years, from the Monsanto data breach in 2011 to the Galil Sewage Corporation's irrigation attack in 2023. Overall, it can be noted that ransomware attacks caused 17 incidents out of 30, and it is the most common form of attack. This review also describes four widely used CSFs and recent AI-based cybersecurity solutions. These recent solutions reported in the survey are applications of AI, blockchain, and FL. By providing this analysis, we aspire that the cybersecurity incidents, along with CSFs and AI solutions, could act as an informative resource to guide respective stakeholders, farmers, and policy makers. Further, this survey introduces a new framework that encompasses the need for AI assurance in the agriculture domain; it also signifies its importance for the validation and verification of AI overall. The FCAI framework proposed in this survey incorporates AI assurance in the AI lifecycle for agricultural applications. Thus, it is crucial for practitioners to consider such approaches and apply AI \& AI assurance as a part of one farm-to-fork agricultural supply chain. This discussion therefore, is aimed at helping organizations and policymakers assess the interrelated effects of cyber threats. Lastly, the cybersecurity incidents in this review may not be exhaustively conclusive, because many may not be publicly available as discussed prior, however, the discussions on the implications and consequences of cyber threats, point to impacts to other critical infrastructures, food security, and the economy as a whole.

\section*{Acknowledgments}
The authors would like to gratefully acknowledge the financial support provided by Deloitte and the Commonwealth Cyber Initiative (CCI). The authors also thank Virginia Tech’s A3 Research Lab (\url{https://ai.bse.vt.edu/}) for their support.

{
\normalsize
\begingroup
\let\itshape\upshape
\bibliography{paper}
\bibliographystyle{unsrtnat}

\endgroup
}

\end{document}